\documentclass[fleqn,usenatbib]{mnras}

\usepackage{newtxtext,newtxmath}
\usepackage[T1]{fontenc}
\usepackage{ae,aecompl}

\usepackage{graphicx}	% Including figure files
\usepackage{amsmath}	% Advanced maths commands
\usepackage{amssymb}	% Extra maths symbols
\usepackage{multirow}
\usepackage{url}
\usepackage{csvsimple,longtable,booktabs}
\usepackage{supertabular}
\usepackage{footmisc}

\title[A Deep Learning Search for Galaxy Clusters]{Deep-CEE I: Fishing for Galaxy Clusters with Deep Neural Nets}

\author[M. C. Chan and J. P. Stott]{
Matthew C. Chan$^{1}$ and John P. Stott$^{1}$
\\
\texttt{E-mails: m.c.chan@lancaster.ac.uk and j.p.stott@lancaster.ac.uk}
\\
$^{1}$Department of Physics, Lancaster University, Lancaster, LA1 4YB, UK
}

\date{Accepted XXX. Received YYY; in original form ZZZ}

\pubyear{2019}

\begin{document}
\label{firstpage}
\pagerange{\pageref{firstpage}--\pageref{lastpage}}
\maketitle

\begin{abstract}
\label{abstract}
We introduce Deep-CEE (\textbf{Deep} Learning for Galaxy \textbf{C}luster \textbf{E}xtraction and \textbf{E}valuation), a proof of concept for a novel deep learning technique, applied directly to wide-field colour imaging to search for galaxy clusters, without the need for photometric catalogues. This technique is complementary to traditional methods and could also be used in combination with them to confirm existing galaxy cluster candidates. We use a state-of-the-art probabilistic algorithm, adapted to localise and classify galaxy clusters from other astronomical objects in SDSS imaging. As there is an abundance of labelled data for galaxy clusters from previous classifications in publicly available catalogues, we do not need to rely on simulated data. This means we keep our training data as realistic as possible, which is advantageous when training a deep learning algorithm. Ultimately, we will apply our model to surveys such as LSST and \textit{Euclid} to probe wider and deeper into unexplored regions of the Universe. This will produce large samples of both high redshift and low mass clusters, which can be utilised to constrain both environment-driven galaxy evolution and cosmology.

\end{abstract}

\begin{keywords}
galaxies: clusters: general -- methods: statistical -- methods: data analysis -- techniques: image processing
\end{keywords}

%%%%%%%%%%%%%%%%%%%%%%%%%%%%%%%%%%%%%%%%%%%%%%%%%%

%%%%%%%%%%%%%%%%% BODY OF PAPER %%%%%%%%%%%%%%%%%%

\section{Introduction}
\label{introduction}

Galaxy clusters are the largest gravitationally bound systems in the Universe. We study galaxy clusters to understand the environmental effects on galaxy evolution and to determine the cosmological parameters that govern the growth of large scale structure in the Universe. In order to do this, large well understood samples of clusters across a range of masses and redshifts are required.

Throughout the 1950s to 1980s the astronomer George Abell created the `Abell catalogue' containing 4073 galaxy clusters \citep{abell_clusters}, which we now refer to as Abell galaxy clusters. George Abell used a magnifying glass to manually examine photographic plates and looked specifically for over-dense regions of galaxies. He could then measure or estimate properties such as distance, richness and galaxy magnitudes for each cluster \citep{abell_review}. That would be the last time that a wide-field cluster search would be conducted manually by eye. 

Since then a variety of techniques have been developed and used to search for galaxy clusters. The primary technique for extracting clusters from imaging data is red sequence fitting (e.g. \citealt{red_sequence_fitting_1}, \citealt{red_sequence_fitting_2}, \citealt{red_sequence_fitting_3}, \citealt{red_sequence_fitting_4}, \citealt{redMapper} and \citealt{red_sequence_fitting_5}). Unlike the Abell method, this technique is applied to photometric catalogue data extracted from imaging, as opposed to the images themselves. It identifies clusters via the distinctive red sequence slopes containing red, passive galaxies found in colour magnitude diagrams. In the charge-coupled device (CCD) era this catalogue-based technique has proven to be an efficient alternative to by-eye searches.

Both X-ray emission (e.g. \citealt{x_ray_emission_1}, \citealt{x_ray_emission_2}, \citealt{x_ray_emission_3}, \citealt{x_ray_emission_4}, \citealt{x_ray_emission_5}, \citealt{x_ray_emission_6}, \citealt{x_ray_emission_7} and \citealt{x_ray_emission_8}) and the Sunyaev-Zeldovich (SZ, \citealt{sz_effect_0}) effect (e.g. \citealt{sz_survey}, \citealt{sz_effect_1}, \citealt{sz_effect_3}, \citealt{sz_effect_4}, \citealt{sz_effect_5}, \citealt{sz_effect_6}, \citealt{sz_effect_7}, \citealt{sz_effect_2} and \citealt{sz_effect_8}) reveal the presence of galaxy clusters through the properties of the hot intracluster medium (ICM). The ICM emits strongly at X-ray wavelengths, so X-ray telescopes such as \textit{XMM-Newton} and \textit{Chandra} can be used to search for clusters, which appear as extended sources. Low energy photons from the Cosmic Microwave Background (CMB) radiation experience Compton scattering when they interact with the high energy electrons of the ICM. This is the SZ effect, in which the presence of a galaxy cluster leaves a shadow on the CMB itself at the galaxy clusters location.

Finally, as clusters are massive structures their presence can also be inferred via weak gravitational lensing (e.g. \citealt{weak_gravitational_lensing_2} and \citealt{weak_gravitational_lensing_1}). Statistical techniques are applied to wide-field galaxy surveys to search for minute signatures of the convergence and alignment of the shears produced by gravitational lensing.  
 
However X-ray, SZ and weak lensing techniques need to optically confirm their candidate clusters as there are contaminants (e.g. active galactic nuclei [AGN] or nearby galaxies) and line-of-sight coincidences (e.g. unrelated low mass groups at different redshifts) that can conspire to give a false detection. This confirmation has to be done by eye, which is time inefficient, or by relying on a red sequence selection again. This can introduce biases or result in an uncertain selection function. Therefore an approach that can produce fast and precise analysis of imaging data would be advantageous to both search for or confirm galaxy clusters.

The Large Synoptic Survey Telescope (LSST, \citealt{LSST_telescope}) is currently under construction in Chile and engineering first-light is expected to be in 2020. LSST will be the deepest wide-field optical survey ever conducted, performing multiple scans of the entire Southern sky over ten years, with an estimated 15\,TB of data generated per night. \textit{Euclid} \citep{euclid} is a wide-field space telescope that is due to commence operation in 2022. It will conduct a weak lensing survey to probe the nature of Dark Matter and Dark Energy, with an estimated 1PB of data generated per year. Data mining techniques such as deep learning will be required to analyse the enormous outputs of these telescopes. LSST and \textit{Euclid} will observe thousands of previously unknown galaxy clusters across a wide range of masses and redshifts but cataloguing them presents a significant challenge. 

During the last two decades computing power has significantly improved \citep{computing_power}, as such deep learning techniques have become an increasingly popular approach to replace repetitive manual tasks. In particular, convolutional neural networks (CNN) have been successful in the field of computer vision, where CNNs are designed to mimic the human brain at learning to perceive objects by activating specific neurons upon visualising distinctive patterns and colours. We can train and utilise CNNs to process high-dimensional features directly from digital images into a meaningful understanding with minimal human input \citep{computer_vision}. \cite{mnist} first introduced a deep learning approach using CNNs to classify uniquely handwritten digits in images from the \textsc{MNIST} dataset achieving error rates of less than 1 per cent. 

Deep learning is very applicable in astronomy due to the abundance of imaging data available from modern telescopes. This makes it preferable when conducting data mining tasks such as classification, regression and reconstruction. These include determining the morphology of galaxies (e.g. \citealt{galaxy_morphology_2}, \citealt{galaxy_morphology_1} and \citealt{galaxy_morphology_3}), identifying gravitational lenses (e.g. \citealt{gravitational_lensing_classification_4}, \citealt{gravitational_lensing_classification_5}, \citealt{gravitational_lensing_classification_6}, \citealt{gravitational_lensing_classification_1}, \citealt{gravitational_lensing_classification_2} and \citealt{gravitational_lensing_classification_3}), predicting galaxy cluster mass (e.g. \citealt{cluster_mass_1}, \citealt{cluster_mass_2}, \citealt{cluster_mass_3}, \citealt{cluster_mass_4} and \citealt{cluster_mass_5}), predicting photometric redshift (e.g. \citealt{photo_redshift_classification_2}, \citealt{photo_redshift_classification_1} and \citealt{photo_redshift_classification_3}), generating synthetic surveys \citep{synthetic_survey_1}, denoising of astronomical images \citep{denoizing_images_1} and astronomical object classification (e.g. \citealt{astro_object_classifier_1} and \citealt{astro_object_classifier_2}). However, a deep learning approach has yet to be developed to detect and determine intrinsic properties of galaxy clusters from wide-field imaging data.

Conventional CNN classifiers are adept at distinguishing learned features in images but rather naive at determining their positions in an image. For this paper we want to apply a deep learning approach that can efficiently localise and classify objects in images. \cite{object_detection} demonstrates a deep learning approach to perform object detection in images by modifying the architecture of CNNs into modules that are specific to classification and localisation tasks, where objects with importance are classed as `foreground' whilst everything else is considered as `background'. 

\textsc{TensorFlow} \citep{tensorflow} is an open source data science library that provide many high level application programming interfaces (API) for machine learning. The object detection API\footnote[1]{The full list of object detection algorithms can be found at: \url{https://github.com/tensorflow/models/blob/master/research/object_detection/g3doc/detection_model_zoo.md} \citep{object_detection_tensorflow}.} \citep{object_detection_tensorflow} contains multiple state-of-the-art deep learning algorithms, which are designed to enhance the speed or accuracy of a model. These include SSD (Single Shot Detection, \citealt{SSD}) and Faster-RCNN (Faster Region-based CNN, \citealt{faster_rcnn_paper}). \cite{object_detection_tensorflow} finds that the Faster-RCNN algorithm returns high precision for predictions on the \textsc{COCO} dataset and is suitable for large input images during training and testing. However, the algorithm can take a long time to train and be slow at generating predictions. Whilst the SSD algorithm is quick to train and produces fast predictions, but the overall precision of predictions is lower compared to Faster-RCNN. We choose the Faster-RCNN algorithm as we prefer accuracy over speed.

We organise this paper in the following format. We split \S \ref{sec:method} into two subsections to outline our methodology. In \S \ref{sec:deep_learning_method} we explain the concept behind the Deep-CEE model\footnote[2]{We use the term `model' to represent the trained `Deep-CEE' Faster-RCNN algorithm.} and in \S \ref{sec:catalogue_and_image_processing} we describe the procedure to create the training and test sets. In \S \ref{results} we analyse the performance of our model with the test set (see \S \ref{Model_Analysis_with_Test_Set}) and we also assess our model on an unseen dataset (see \S \ref{Comparison_To_redMapper_galaxy_clusters}). In \S \ref{Discussion} we discuss the limitations and future applications of our model. Finally, in \S \ref{Conclusion} we summarise this paper.

Throughout this paper, we adopt a $\Lambda$CDM cosmology with $H_{0} = 71 \ \text{km} \ \text{s}^{-1} \ \text{Mpc}^{-1}$, $\Omega_{m} = 0.27$ and $\Omega_{\Lambda} = 0.73$.

\section{Method}
\label{sec:method}

\subsection{Deep Learning Method}
\label{sec:deep_learning_method} 

\begin{figure*}
	\includegraphics[width=\linewidth]{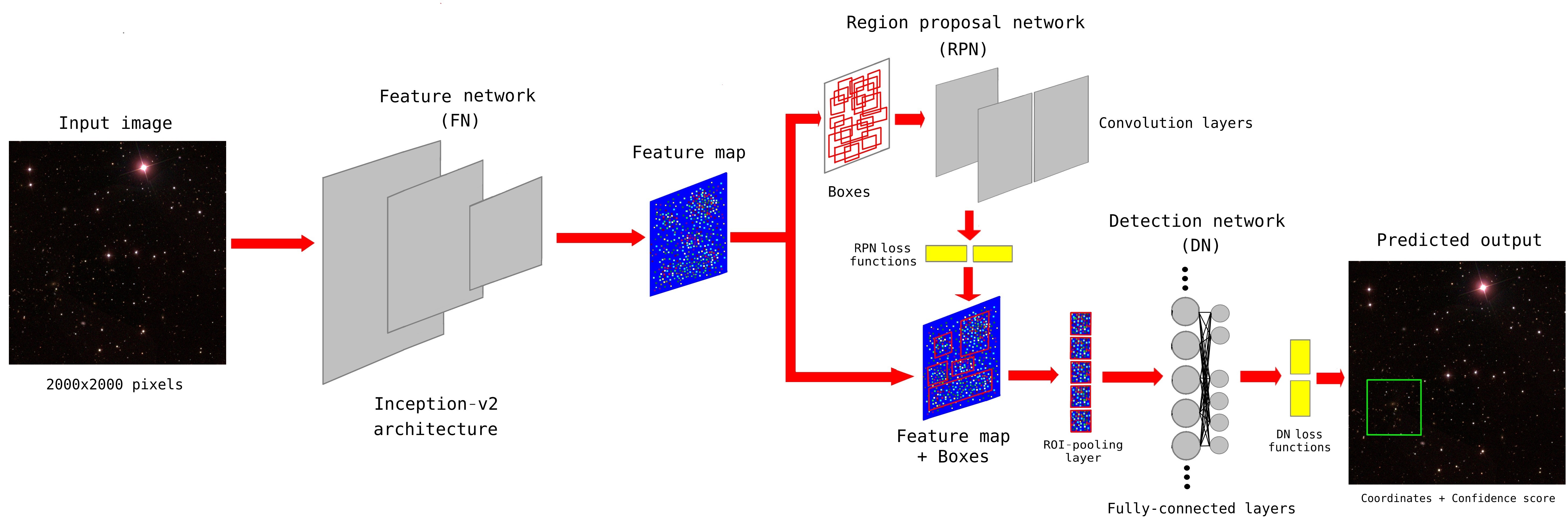}
    \caption{High-level overview of the architecture for the Faster-RCNN algorithm which contains the Feature Network, Region Proposal Network and Detection Network. The output from each network is used as the input for the next network. The algorithm's architecture is similar to the system demonstrated in Figure 2 from \protect\cite{faster_rcnn_paper}. For simplicity, the \textsc{Inception-v2} architecture is not displayed fully. It should be noted that `$Mixed\_4e$' is used as the final layer of the Feature Network \citep{object_detection_tensorflow}. The full details of the \textsc{Inception-v2} architecture can be found in \protect\cite{faster_rcnn_architectures}. The RPN and the DN loss functions are only active during the training phase. A softmax function \protect\citep{softmax} is used with the RPN and DN loss functions for classification of proposed boxes.}
    \label{fig:faster_rcnn_model}
\end{figure*}

We use a supervised learning approach \citep{supervised_learning} to train the Faster-RCNN algorithm by providing it labelled data. The architecture of the algorithm can be seen in Figure \ref{fig:faster_rcnn_model}. It is comprised of three different individual networks that work collectively during the training phase. These three networks are called the Feature Network (FN), Region Proposal Network (RPN) and Detection Network (DN). The convolution layers in the RPN and the fully-connected (FC) layers in the DN require training as the other layers in the RPN and the DN have no trainable parameters. To train our model, we use a joint training \citep{faster_rcnn_paper} approach, which means we allow the outputs from all the networks to be generated before all trainable layers are updated. Throughout this section we adopt a similar methodology and hyper-parameters as described in \cite{faster_rcnn_paper}. We set a learning rate of 0.0002, momentum of 0.9, gradient clipping threshold of 10 and a mini-batch size of one\footnote[3]{The definitions for each hyper-parameter is beyond the scope of this paper but is explained in \cite{hyper_parameters}.}. We perform random initialisation of the weights from a zero-mean truncated Gaussian distribution with a standard deviation of 0.01 for the RPN and we apply variance scaling initialisation \citep{xavier_initialisation} from a uniform distribution for the weights in the DN. We also initialise bias values for the trainable layers in the RPN and DN to be zero.

\subsubsection{Feature Network}
\label{feature_network}

The FN is found at the beginning of the Faster-RCNN algorithm and takes an image as the input. We apply transfer learning \citep{transfer_learning} by using a pre-trained CNN called \textsc{Inception-v2} \citep{inception_module_v2} as the architecture of the FN. \textsc{Inception-v2} consists of convolution, ReLU (rectified linear unit) activation and pooling layers\footnote[4]{The fully-connected layers and softmax function} in \textsc{Inception-v2} are not included for the FN since we do not want to perform classification or regression in this network. \citep{deep_learning}. The convolution and ReLU activation layers are responsible for extracting non-linear features from an image such as straight lines, edges and curves. The pooling layers then down-sample the convolved image to form a feature map. The reason we choose the \textsc{Inception-v2} architecture as opposed to other architectures such as \textsc{VGG16} \citep{vgg16} and \textsc{AlexNet} \citep{alexnet} is that it has been designed to reduce the number of parameters needed in the network. This means less computational power is required to train the algorithm, while still retaining high accuracy. \textsc{Inception-v2} has been pre-trained to recognise objects from the \textsc{COCO} (The Common Objects in Context) dataset \citep{COCO_dataset}, which contains images of commonly found objects in daily life such as vehicles and animals. This means we do not have to fully retrain the weights and biases in the network since they are sufficiently optimised at finding generic structures, as retraining every single weight and bias from scratch is computationally expensive. Furthermore we do not alter the architecture of \textsc{Inception-v2}, as it has already been tuned for object detection. 

\subsubsection{Region Proposal Network}
\label{region_proposal_network}

The RPN is found after the FN and consists of a shallow architecture of three convolution layers with a ReLU activation layer specific only to the first convolution layer, see Figure \ref{fig:faster_rcnn_model}. The weights and biases in the the first convolution layer are shared for classification and localisation tasks whilst the remaining convolution layers are separated into two parallel convolution layers, with independent weights and biases for each task. The RPN takes the feature map output from the FN as its input. The role of the RPN is to generate possible positions at which an object could be located within an image. In the first convolution layer, a $3 \times 3$ pixel sliding window is used with zero-padding\footnote[5]{Additional layers of pixels are added around the edge of an image with values of zero. This helps to preserve the dimensions of the input as it passes through the layer.} and a pixel stride of one, which translates to every sixteenth pixel in the original image. At the centre of each sliding window an `anchor' is placed. Each anchor has a set number of different sized boxes generated around it. The dimensions and number of boxes is dependent on the scaling and aspect ratios. We set scaling ratios of 0.25, 0.5, 1.0 and 2.0 and aspect ratios of 0.5, 1.0 and 2.0. We choose these values to reflect the dimensions of the ground truth boxes in all of the images. A scaling ratio of 1.0 relates to a box of $256 \times 256$ pixels in the original image, such that setting other values for the scaling ratio would generate additional larger or smaller boxes at each anchor. The aspect ratio produces boxes around each anchor that have adjusted widths and heights with respect to each scaling ratio. This means at every anchor there are twelve boxes of different sizes. In the final convolution layers, a $1 \times 1$ pixel sliding window is used with no-padding\footnote[6]{No additional pixels are added around the edge of an image.} and a pixel stride of one. This ensures fixed dimensions for the output of this layer.

Any box that overlaps by more than 70 per cent with the ground truth box is assigned as a positive `foreground' label or otherwise a negative `background' label if less than 30 per cent, whilst boxes between these overlap thresholds are ignored. From which 128 positive labelled boxes and 128 negative labelled boxes are randomly chosen for each image to update the weights and biases, so that the RPN learns to distinguish important objects as `foreground' and irrelevant objects as `background'. If there are fewer than 128 positive labelled boxes in an image then additional negative labelled boxes with the next highest percentage overlapping are chosen to represent positive labels. Therefore the RPN will learn two outputs: whether a box is likely to contain a ground truth object and whether a box is not likely to contain a ground truth object, based on percentage overlap with the ground truth box. During the testing phase, if a box has a high probability of containing an object then this box will be passed onto the next stage in the Faster-RCNN algorithm. However, if a box has a high probability of not containing an object then the box is disregarded. Back-propogation (BP, \cite{backpropogation}) and stochastic gradient descent (SGD, \cite{stochastic_gradient_descent}) is used to train the weights and biases in RPN\footnote[7]{Back-propogation with respect to box coordinates is disabled as it causes the training to be unstable \citep{object_detection_tensorflow}\label{first_footnote}.}.

Two additional steps are applied to limit the number of boxes for faster computation. Firstly, any box which extends outside the image borders are disregarded after the boxes are generated. Non-Maximum Suppression (NMS, \cite{non_maximum_suppression}) is used on the boxes after the losses are calculated. NMS keeps the highest overlapping box with the ground truth box and disregards any remaining boxes that overlap by more than 70 per cent with this box. These steps are repeated on the remaining boxes, such that the next box with the highest overlap is kept and any other box with more than 70 per cent overlap with this box are also disregarded. This procedure continues until there are fewer than 300 boxes for each image.

We utilise two loss functions (categorical cross-entropy loss and smooth L1-loss) to calculate prediction errors in the RPN, these two loss functions are associated to separate convolution layers in the final layer of the RPN. The categorical cross-entropy loss layer \citep{cross_entropy_loss} creates a probability distribution for all the proposed boxes, where the sum of all probabilities equals one. This function is described in Equation \ref{eq:cross_entropy}:

\begin{equation}
\begin{split}
    L_{cls}(p_{i},p_{i}^*) = -p_{i}^* log(pi_{i}) - (1-p_{i}^*)log(1-p_{i}),
	\label{eq:cross_entropy}
\end{split}
\end{equation}

where ${p}_{i}$ is the predicted probability of a box and ${p}_{i}^*$ is zero or one depending on whether the ground truth box is correctly classified. The categorical cross-entropy loss function calculates the objectness error for boxes being predicted as `foreground' and `background'. 

The smooth L1-loss layer \citep{fast_rcnn} only considers positive labelled boxes in the training phase. It takes into account the distance between the centre coordinates of the ground truth box and predicted boxes, and also the difference in size of the boxes compared to the ground truth box. This function is seen in Equation \ref{eq:smooth_loss}:

\begin{equation}
    L_{reg}(x) = \Big\{
    \begin{tabular}{cc}
    $0.5x^2$, & \text{if} $|x| < 1$, \\
    $|{x}| - 0.5$, & \text{otherwise},
    \end{tabular}.
    \label{eq:smooth_loss}
\end{equation}

where $x$ = $({t}_{i}-{t}_{i}^*)$ is the error between the ground truth and the predicted boxes. The smooth L1-loss function penalises localisation error by taking the absolute value (behaves like L1-loss) for large errors and the square value (behaves like L2-loss) for small errors \citep{l1_l2_loss}. This encourages stable regularisation of the weights and biases during training.

The proposed boxes from the RPN are merged with the feature maps from the FN, so that each box is overlaid on an `object' in a feature map. An ROI-pooling (region-of-interest) layer \citep{roi_warping} is then applied, which divides each box into the same number of sections\footnote[8]{Tensorflow's `$crop\_and\_resize$' operation is used to replicate the ROI-pooling layer in \cite{faster_rcnn_paper} \citep{object_detection_tensorflow}.}. The largest value in each section is extracted to generate new cropped feature maps associated to each box from the previous feature map. ROI-pooling speeds up computation later on in the Faster-RCNN algorithm, as having fixed sized feature maps leads to faster convergence \citep{fast_rcnn}.

\subsubsection{Detection Network}
\label{detection_network}

The DN is found after the ROI-pooling layer at the end of the Faster-RCNN algorithm. The DN is composed of FC layers, see Figure \ref{fig:faster_rcnn_model}. The purpose of an FC layer is to combine all the outputs from the previous layer, this allows for the algorithm to make decisions. The FC layers are run in parallel, such that the weights and biases are split between classification and localisation. One of these two FC layers consists of 2 neurons to categorise the outputs for classification, the other FC layer consists of 4 neurons to predict the properties for box regression. Similar to the procedure for RPN, 16 positive and 48 negative labelled boxes are randomly chosen in each image to train the weights and biases in the DN, where boxes that overlap by more than 50 per cent with the ground truth box are assigned as positive `foreground' labels or otherwise negative `background' labels. Additional negative boxes with the next highest overlap are assigned as positive labels if there are fewer than 16 positive labelled boxes. At the end of the DN, another categorical cross-entropy loss and smooth-L1 loss layer is used to calculate the classification and localisation errors, where each loss function is also associated to its own FC layer in the DN. NMS is applied again using a 60 per cent threshold to reduce the number of overlapping boxes until fewer than 100 boxes remain. The weights and biases in the FC layers are also trained via BP and SGD\footref{first_footnote}. Classification loss is measured by comparing the cropped feature maps of each box with the ground truth box. We determine box regression loss by calculating the difference between the pixel coordinates, height and width of the positive labelled boxes with the ground truth box.

Finally, the loss functions of the RPN and the DN are combined into one multi-tasking loss function (\citealt{faster_rcnn_paper} and \citealt{object_detection_tensorflow}) to train the algorithm. It takes into account the classification loss and box regression loss, by comparing all of the predicted properties for the boxes with the ground truth box in each image. Therefore the total loss of the algorithm is the weighted sum of the RPN and DN losses. The multi-tasking loss function is described in Equation \ref{eq:total_loss}:

\begin{equation}
\begin{split}
    L(\{p_{i}\},\{t_{i}\}) =\alpha \frac{1}{N}\sum_i L_{cls}(p_{i},p_{i}^*) \\ + \beta \frac{1}{N}\sum_i p_{i}^* L_{reg}(t_{i},t_{i}^*),
	\label{eq:total_loss}
\end{split}
\end{equation}

where $i$ is the box index number in each mini-batch, $p_{i}$ is the predicted probability of a labelled box, ${t}_{i}$ represents the height, width, center x and y coordinates of a labelled box, $\alpha$ and $\beta$ are balancing weights for the classification and regression terms, $p_{i}^*$ signifies positive/negative box labels, $t_{i}^*$ is the ground truth box properties and $N$ is the number of anchors/proposals selected to compute the loss function per mini-batch. ${L}_{cls}$ is classification loss and ${p}_{i}^* {L}_{reg}$ is box regression loss for only positive labelled boxes. 

\subsection{Catalogue and Image Pre-Processing}
\label{sec:catalogue_and_image_processing} 

\cite{clusters_catalogue} applied the Friends-of-Friends cluster detection algorithm \citep{friend_of_friend} on photometric data taken from the Sloan Digital Sky Survey III (SDSS-III, \citealt{sdss_III}) Data Release 8 (DR8, \citealt{sdss_dr8}). \cite{clusters_catalogue} identified 132,684 galaxy clusters in the redshift range $0.05 \leq z < 0.8$. The resultant catalogue has a completeness ratio of > 95 per cent for detecting galaxy clusters with mass greater than $ 1.0 \times 10^{14} \ \text{M}_{\odot}$ inside $\text{R}_{200}$\footnote[9]{$\text{R}_{200}$ is the radii at which the mean density of the galaxy cluster is 200 times greater than the critical density of the Universe \citep{r200_radius}.} and in the redshift range of $0.05 \leq z < 0.42$. We use the Abell galaxy clusters identified in the \cite{clusters_catalogue} catalogue, to obtain the labelled data needed to create the training set. We choose the Abell galaxy clusters because our technique uses visual inspection of images in a similar manner to that performed by George Abell and is therefore appropriate to this proof of concept paper.

\begin{figure}
	\includegraphics[width=\linewidth]{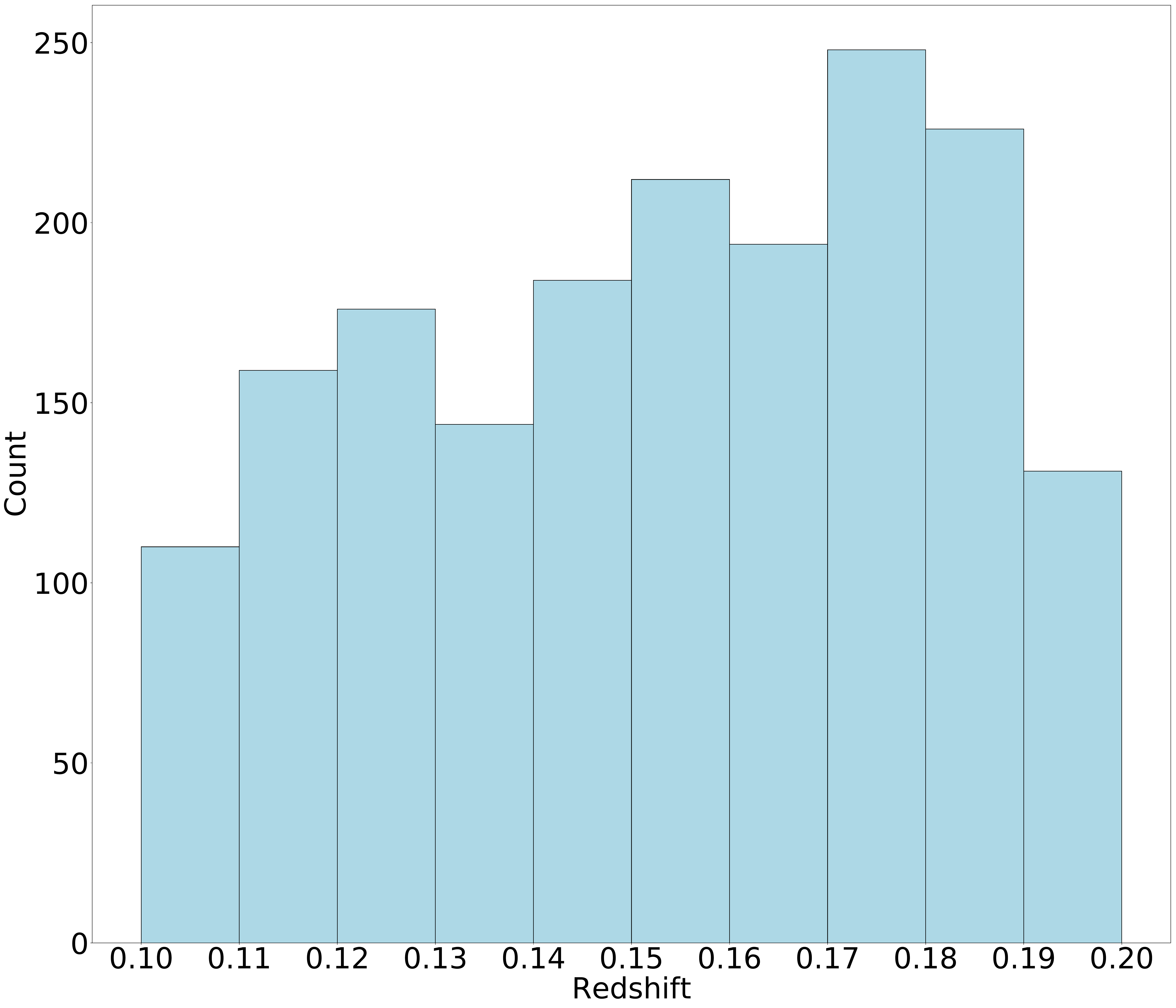}
	\includegraphics[width=\linewidth]{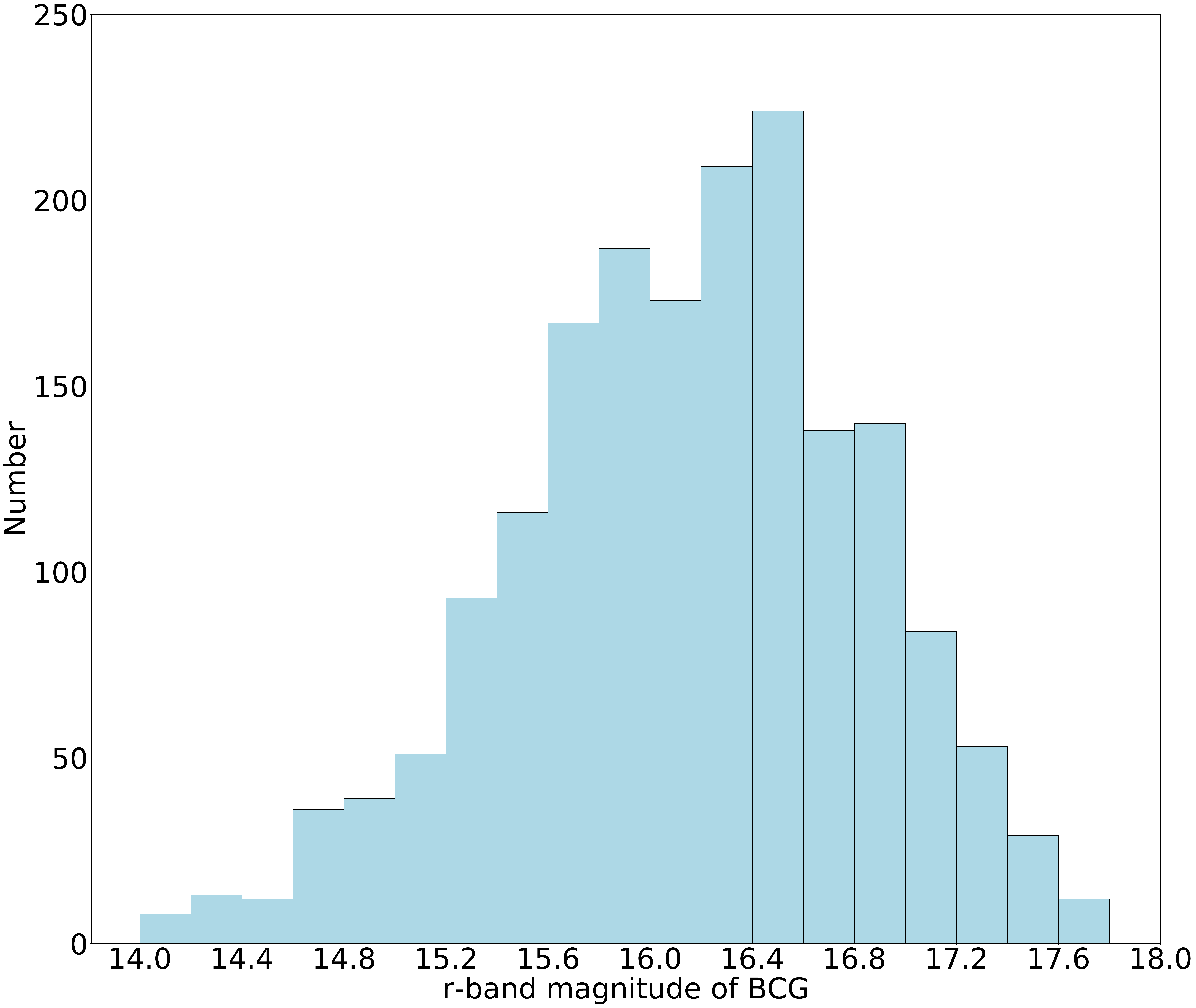}
	\includegraphics[width=\linewidth]{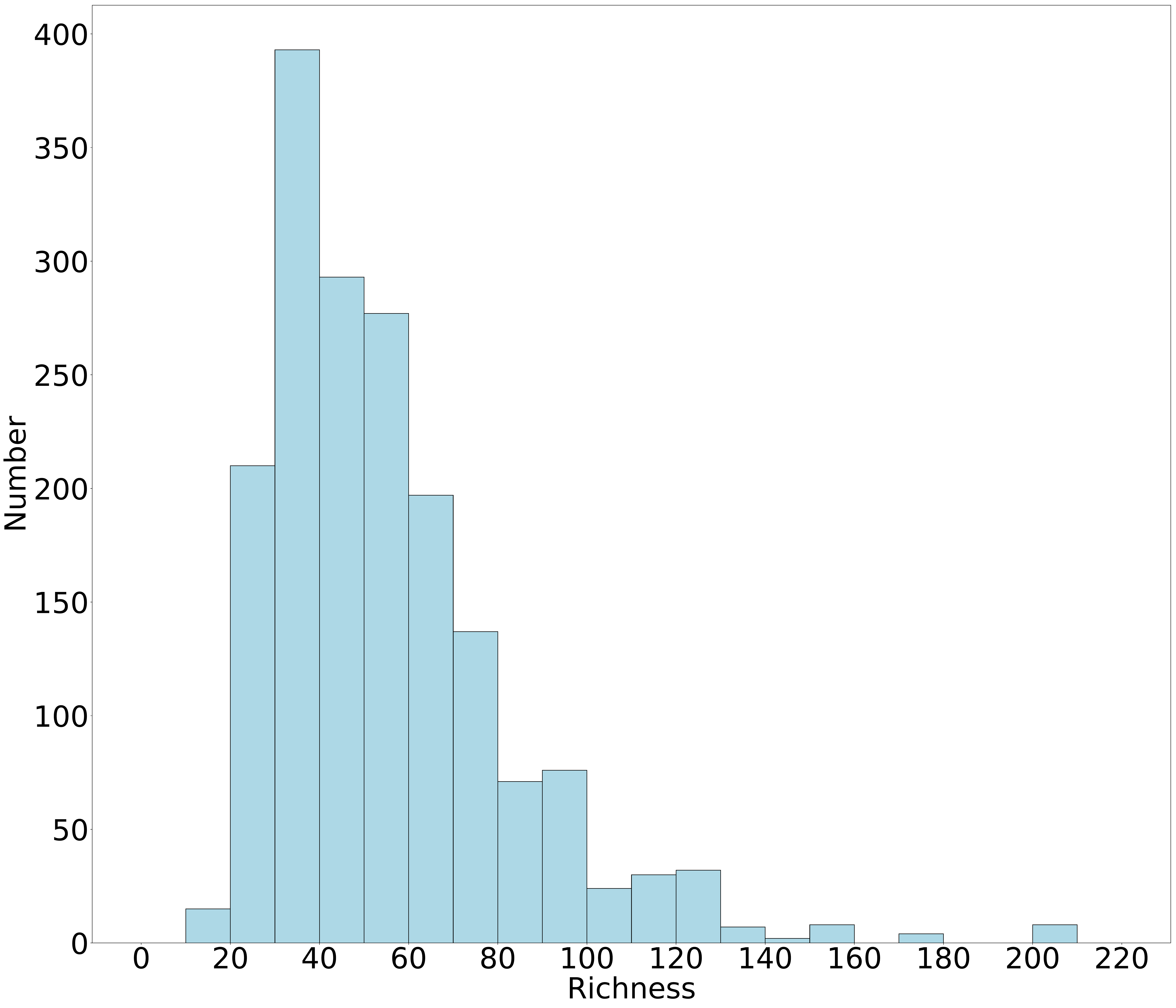}
    \caption{The distributions of the properties for all of the galaxy clusters in the training set. The histograms display the photometric redshift, $r$-band magnitude of the BCG and richness (from top to bottom).}
    \label{fig:train_properties}
\end{figure}

We do not train the algorithm on the entire \cite{clusters_catalogue} catalogue, as this is a pilot study to test the performance of the Faster-RCNN algorithm at identifying galaxy clusters from a sample set. We limit the photometric redshift range of galaxy clusters to $0.1 < z < 0.2$, as we want to maximise the signal-to-noise available and avoid nearby galaxy clusters that could fill the field of view. See Figure \ref{fig:train_properties} for the photometric redshift distribution in the training set. We set a threshold of $20 \ge $ galaxy members inside a $R_{200}$ radius, as poorly populated galaxy clusters may have a lower signal-to-noise. Applying these constraints results in a sample set of 497 Abell galaxy clusters. Figure \ref{fig:train_properties} shows the richness distribution of the galaxy clusters in the training set within $\text{R}_{200}$. Richness is defined by Equation \ref{eq:richness}:

\begin{equation}
    R_{L^*}=\frac{L_{200}}{L^{*}},
	\label{eq:richness}
\end{equation}

where $R_{L^*}$ is the galaxy cluster richness, $L_{200}$ is the total $r$-band luminosity within $R_{200}$ and $L^{*}$ is the typical luminosity of galaxies in the $r$-band \citep{clusters_catalogue}.

The brightest cluster galaxy (BCG) is a giant elliptical galaxy that is usually located in the vicinity of the spatial and kinematic centre of a galaxy cluster \citep{BCG_cluster}. We convert the right ascension (RA) and declination (Dec) of the BCG in each image to pixel coordinates. We adopt these pixel coordinates as the centre coordinates for the ground truth boxes in both training and test sets. Figure \ref{fig:train_properties} shows the distribution of the $r$-band magnitudes for the BCGs in the training set. We also restrict Dec to greater than 0 degrees to reduce the amount of data near the galactic plane, due to a higher concentration of stars that may introduce significant foreground contamination.

\begin{figure}
	\includegraphics[width=\linewidth]{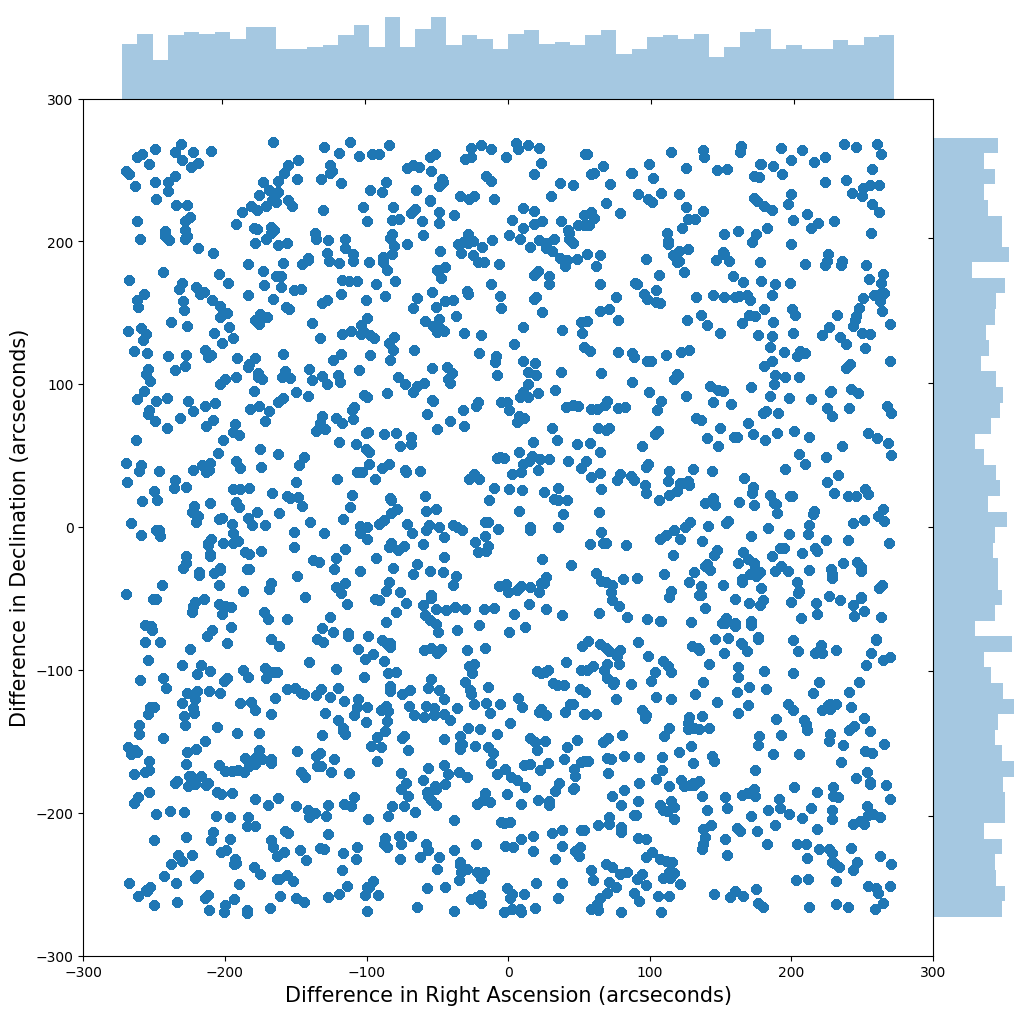}
	\includegraphics[width=\linewidth]{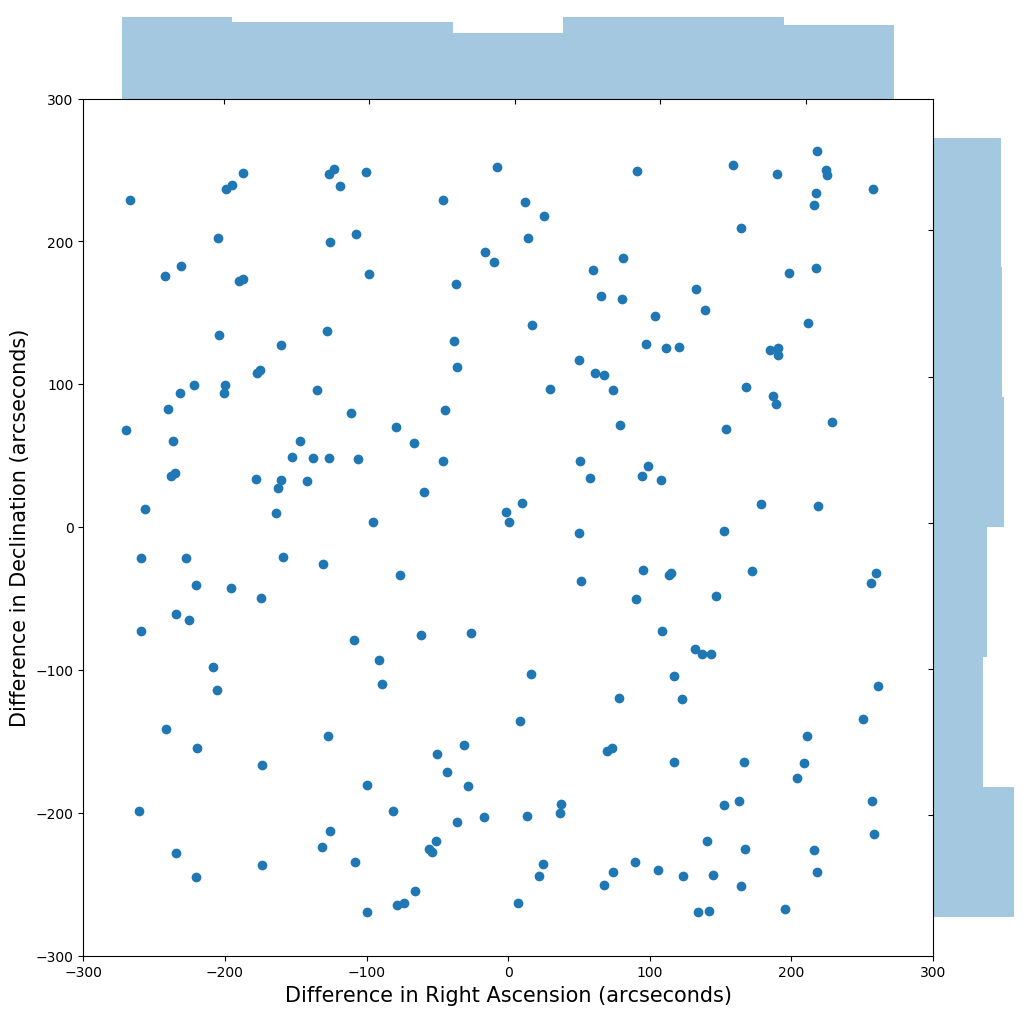}
    \caption{The distribution of image positions for the galaxy clusters in the training set (top) and test set (bottom). The positions are determined by calculating the difference in arcseconds between the coordinate offset and the true coordinates of the galaxy cluster at its respective photometric redshift, where the coordinate offsets are sampled from a uniform distribution.}
    \label{fig:image_offset}
\end{figure}

The imaging camera on the SDSS telescope has a pixel size scaling of $0.396$ \text{arcsec} $\text{pixel}^{-1}$. The SDSS telescope has five broadband imaging filters, referred to as \textit{u, g, r, i, z} covering a wavelength range of 3543 to 9134 \AA \ \citep{sdss_III}. We used the \textit{i, r, g} filters but not the \textit{u} and \textit{z} filters as the sensitivity of the SDSS telescope is poorer at these wavelengths. Each image in the training set contains one Abell galaxy cluster labelled as a ground truth. We fix each image size to $2000 \times 2000$ pixels (approximately $1443 \times 1443$ kpc at redshift $z=0.1$ and $2588 \times 2588$ kpc at redshift $z=0.2$) to capture the wider context but we lower the resolution of the images during training to a fixed dimension of $1000 \times 1000$ pixels for computational efficiency. We apply a random offset from a uniform distribution to the input coordinates since we do not want the algorithm to be biased towards specific positions in an image. This offset results in a uniform spread of the positions of the galaxy clusters in all images, where a galaxy cluster could be found anywhere within 270 arcseconds from the x and y plane of the image centre. See Figure \ref{fig:image_offset} for the distribution of the positions in the training and test sets.

To make colour images we ensure that the individual images taken from the publically available SDSS-III Data Release 9 (DR9, \citealt{sdss_dr9})\footnote[10]{The imaging data for SDSS-III DR9 can be found via NASA's SkyView (\url{http://skyview.gsfc.nasa.gov}) online database \citep{skyview}.} for the \textit{i, r, g} filters are set to the same scaling and aspect ratios. We stack the three filter images to RGB channels and apply a non-linear transformation to `stretch' each image channel. We experimented with linear, square root and logarithmic transformations. These transformations adjust the contrast of the image, as we define lower and upper flux limits by mapping the image onto a luminosity scale. We do this to reduce background noise, dim extremely bright objects and make stars and galaxies easily distinguishable. We find the square root function to be better for visualizing galaxy structure and colour. This benefits the algorithm by decreasing the learning complexity of the features.

\begin{figure*}
	\includegraphics[width=\linewidth]{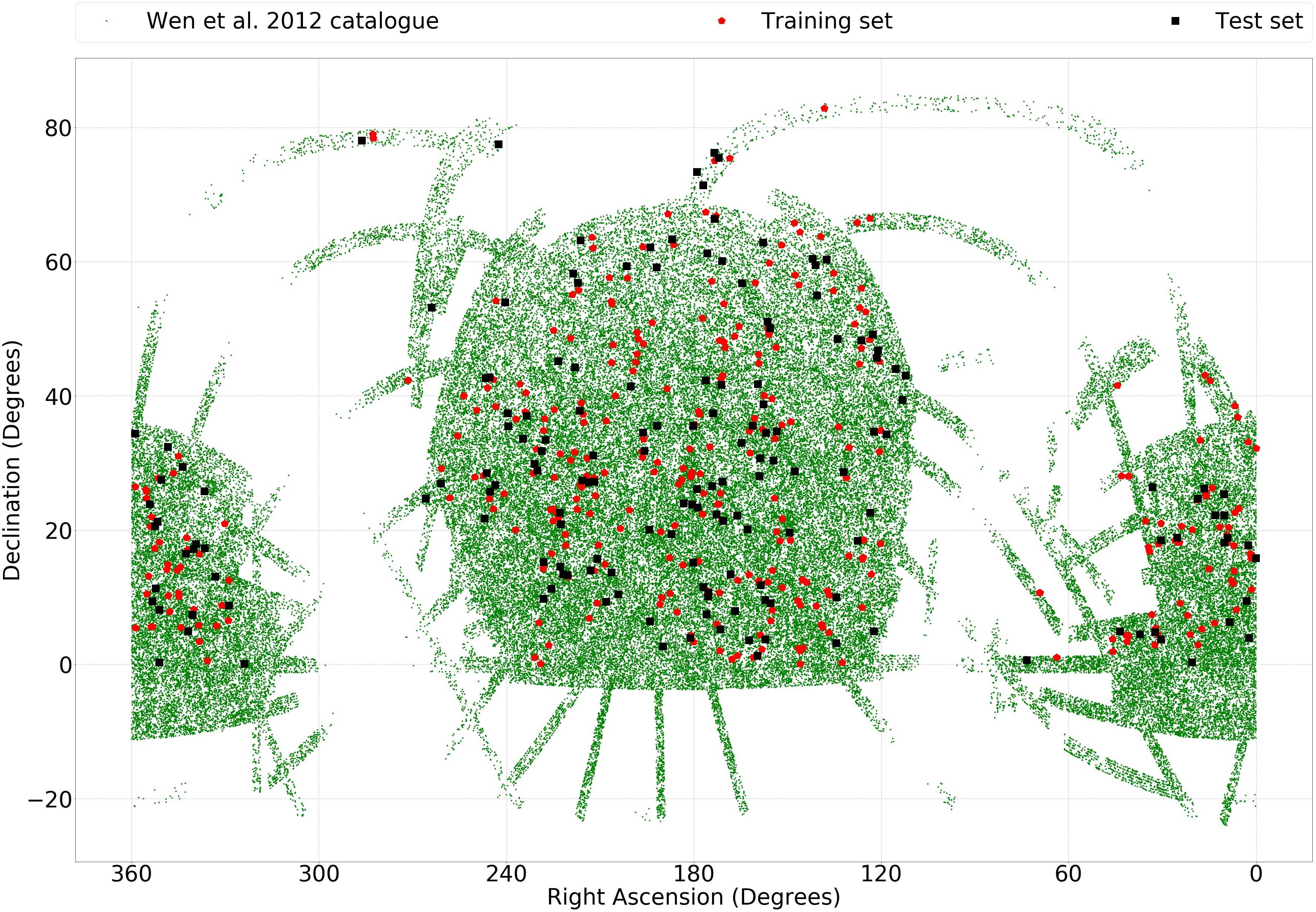}
    \caption{A map of astronomical coordinates using the J2000 epoch system of the galaxy clusters in the training set, test set and \protect\cite{clusters_catalogue} catalogue.}
    \label{fig:RA_and_DEC_map}
\end{figure*}

We increase the amount of variance in the sample set by applying image augmentation techniques. In \cite{faster_rcnn_paper}, it is stated that one of the properties of the Faster-RCNN algorithm is translational invariance, which means the algorithm is robust at learning translated objects. We train the algorithm to recognise that an object could appear at any location in an image. Since our method applies a random offset to the input coordinates via translation we augment the sample set three additional times, which boosts the sample size to 1988. This also introduces additional negative boxes for the algorithm to learn, as each image contains a different background. We randomly shuffle the sample set and perform simple random sampling to split the sample set into a training and test set, which are approximated representations of the full population. The training set is made up of ${\sim}90$ per cent of the sample set consisting of 1784 labelled galaxy clusters and the test set is made up of the remaining ${\sim}10$ per cent consisting of 204 labelled galaxy clusters. Figure \ref{fig:RA_and_DEC_map} displays the astronomical coordinates for the galaxy clusters in the training and test sets compared to all the galaxy clusters in the \cite{clusters_catalogue} catalogue. We also horizontally flip the images in the training set, where each image has a 50 per cent chance of being flipped. This approach can double the size of the training set to 3568 if all images are flipped once but this does not affect the size of the test set. Since galaxy clusters can be observed from any orientation we find these augmentation techniques to be appropriate during training.

\section{Results}
\label{results}

\subsection{Model Analysis with Test Set}
\label{Model_Analysis_with_Test_Set}

\begin{figure}
	\includegraphics[width=\linewidth]{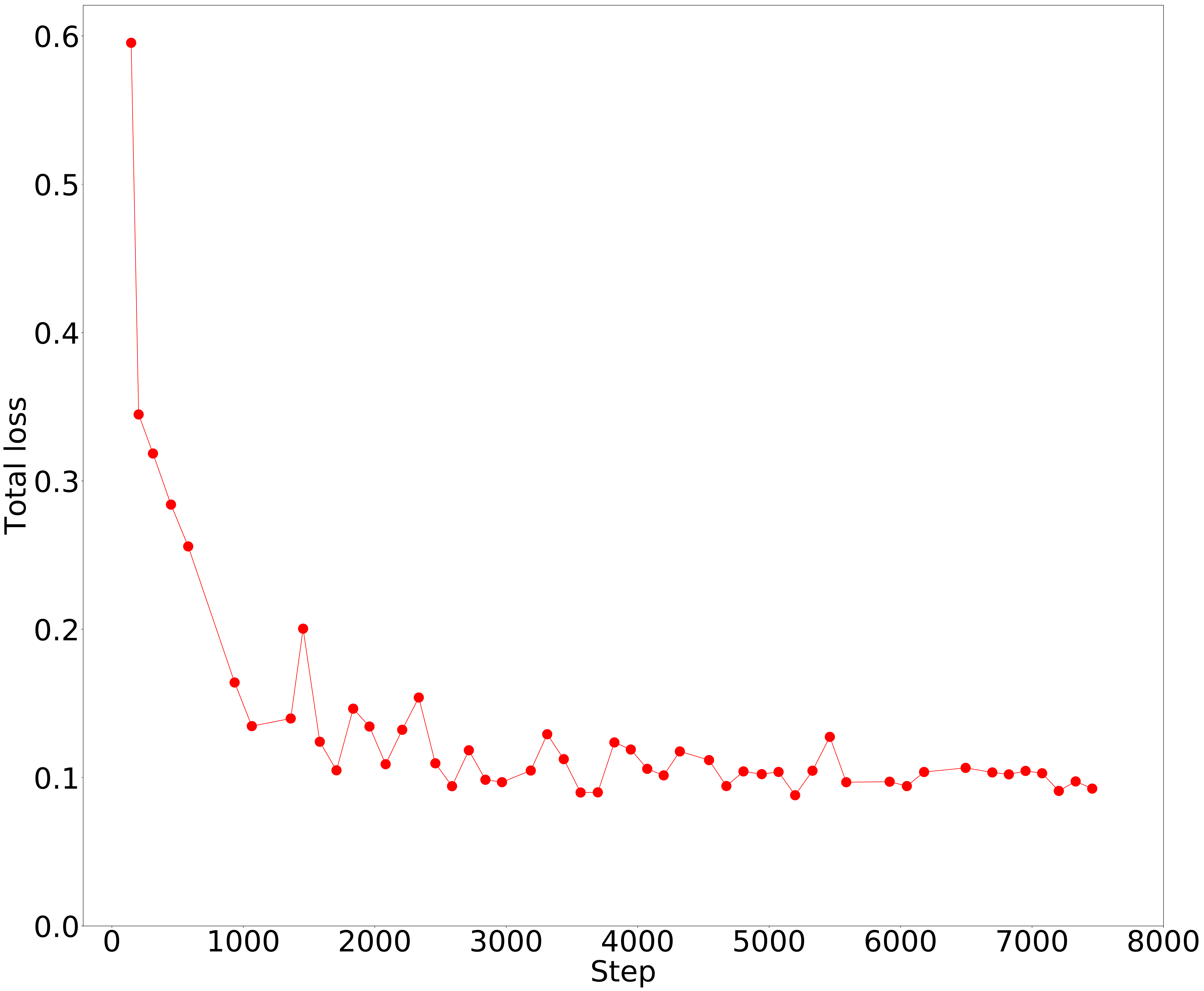}
    \caption{The total loss (see Equation \ref{eq:total_loss}) considers the errors from the RPN and the DN during training. We stop the the training of the model after evaluation at 7458 steps as the loss stabilises. Each point represents the total loss recorded at different step intervals. The values of these points can be found in Table \ref{tab:all_losses_table}.}
    \label{fig:total_loss}
\end{figure}

We train our model with graphics processing unit (GPU) support for a maximum of 25,000 steps to ensure the algorithm has enough training time to sufficiently minimise prediction errors. The number of steps is a tunable hyper-parameter that can shorten or extend the run-time of training a model. In Figure \ref{fig:total_loss}, we find that the algorithm generalises well as the total loss stabilises at approximately 3,000 steps, where a step represents one iteration per mini-batch from the dataset through the algorithm. For a competent model, the total loss should not fluctuate significantly during training. 

\begin{figure*}
    \includegraphics[width=0.49\linewidth]{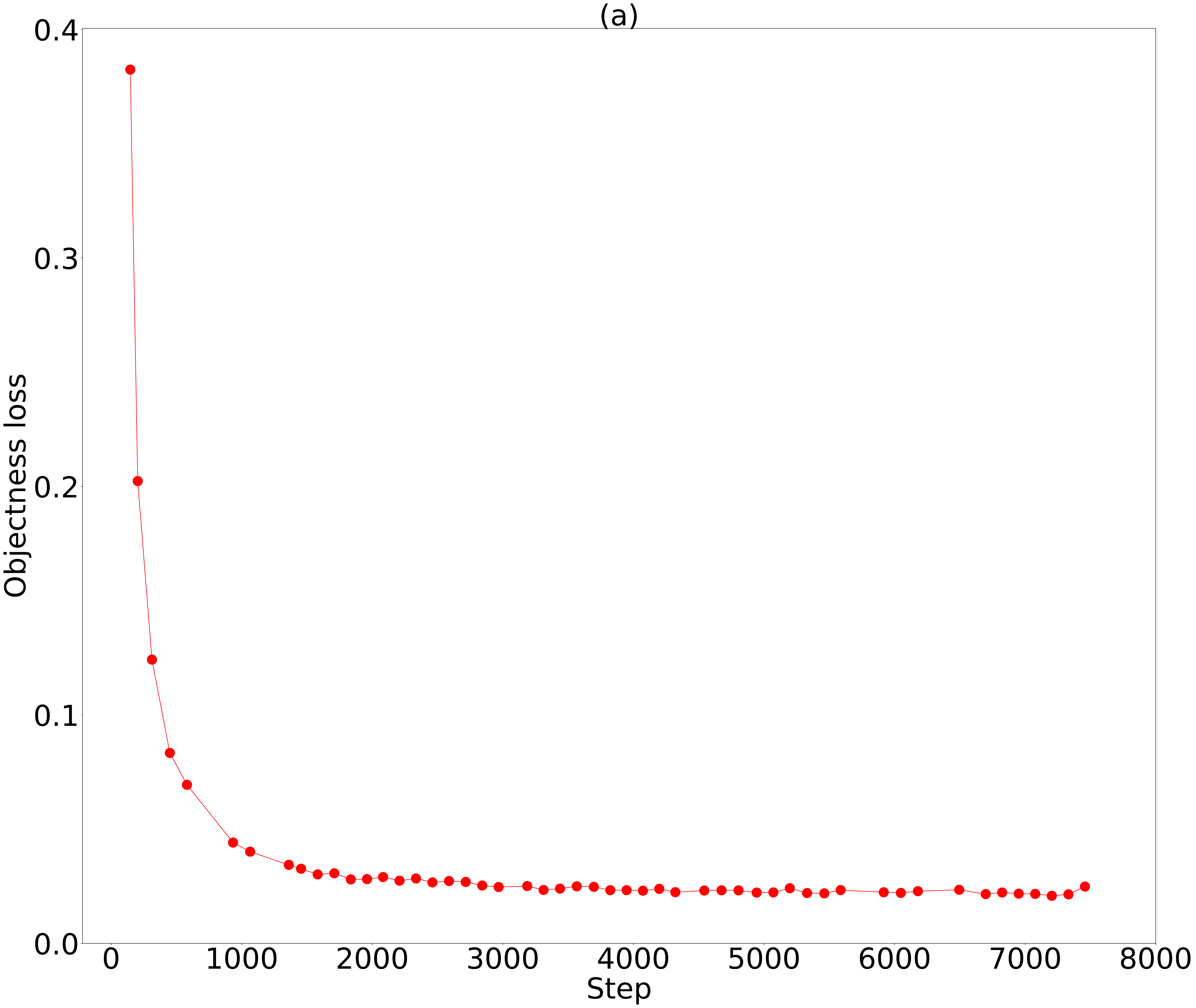}
    \includegraphics[width=0.49\linewidth]{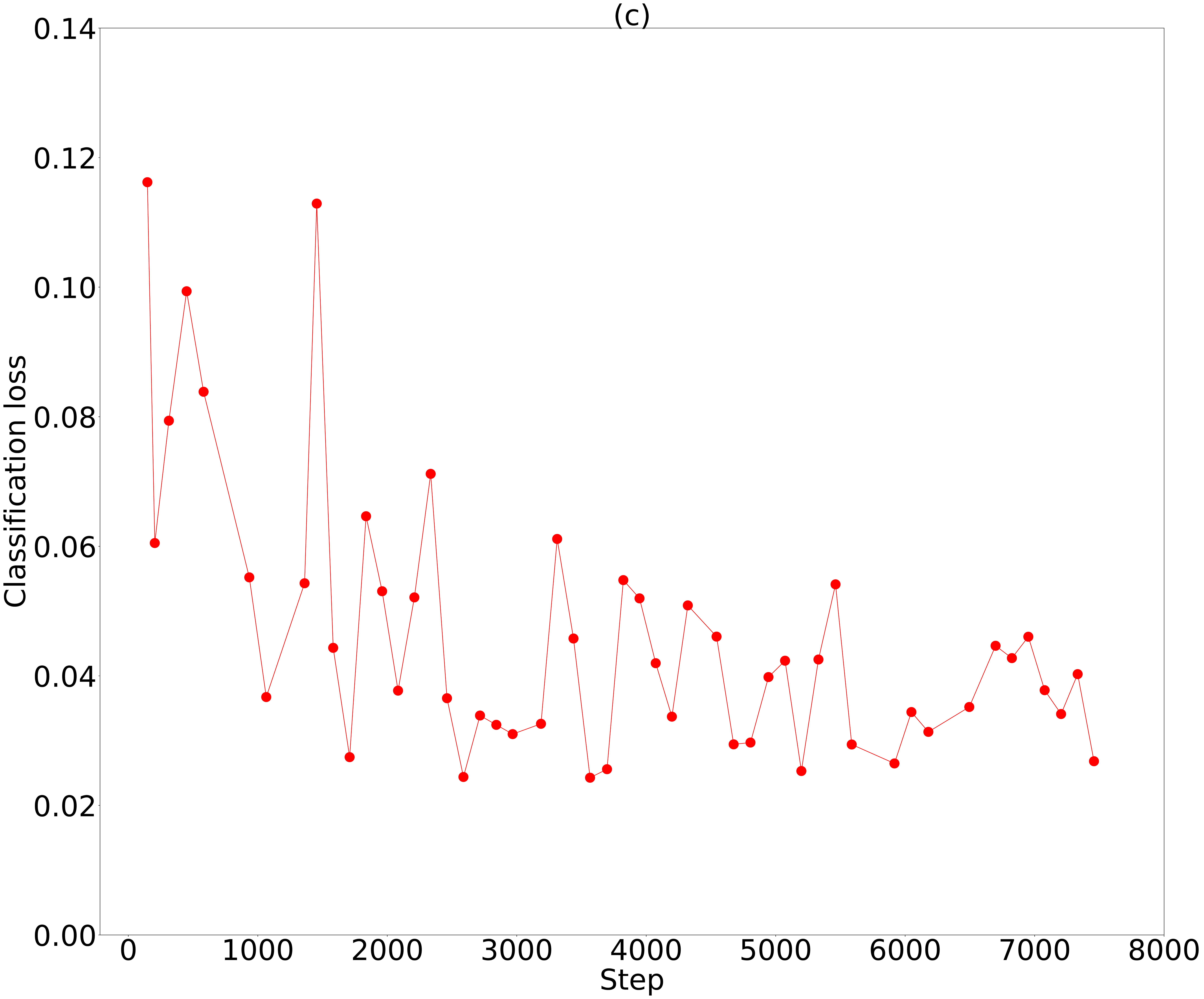}
    \includegraphics[width=0.49\linewidth]{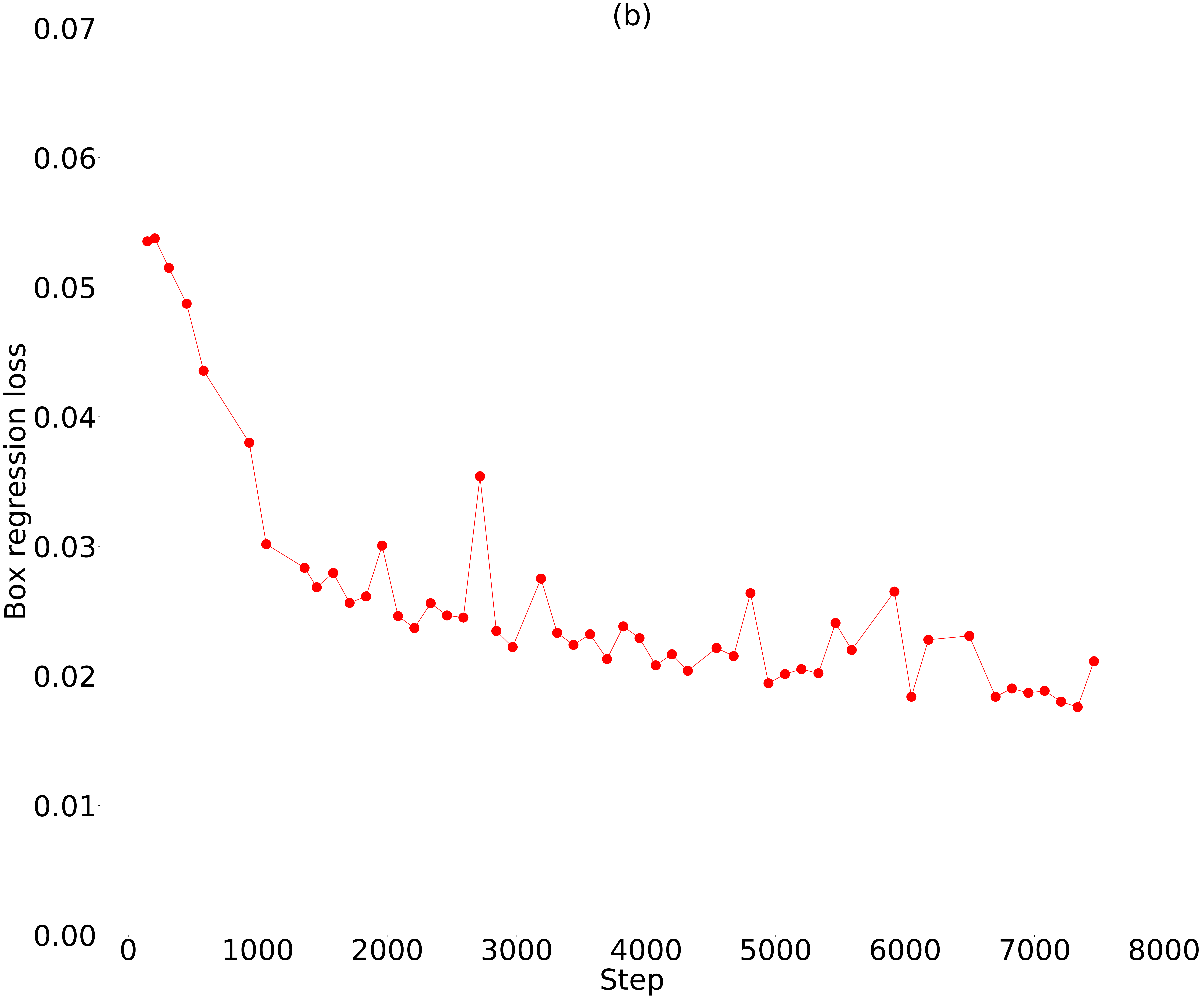}
    \includegraphics[width=0.49\linewidth]{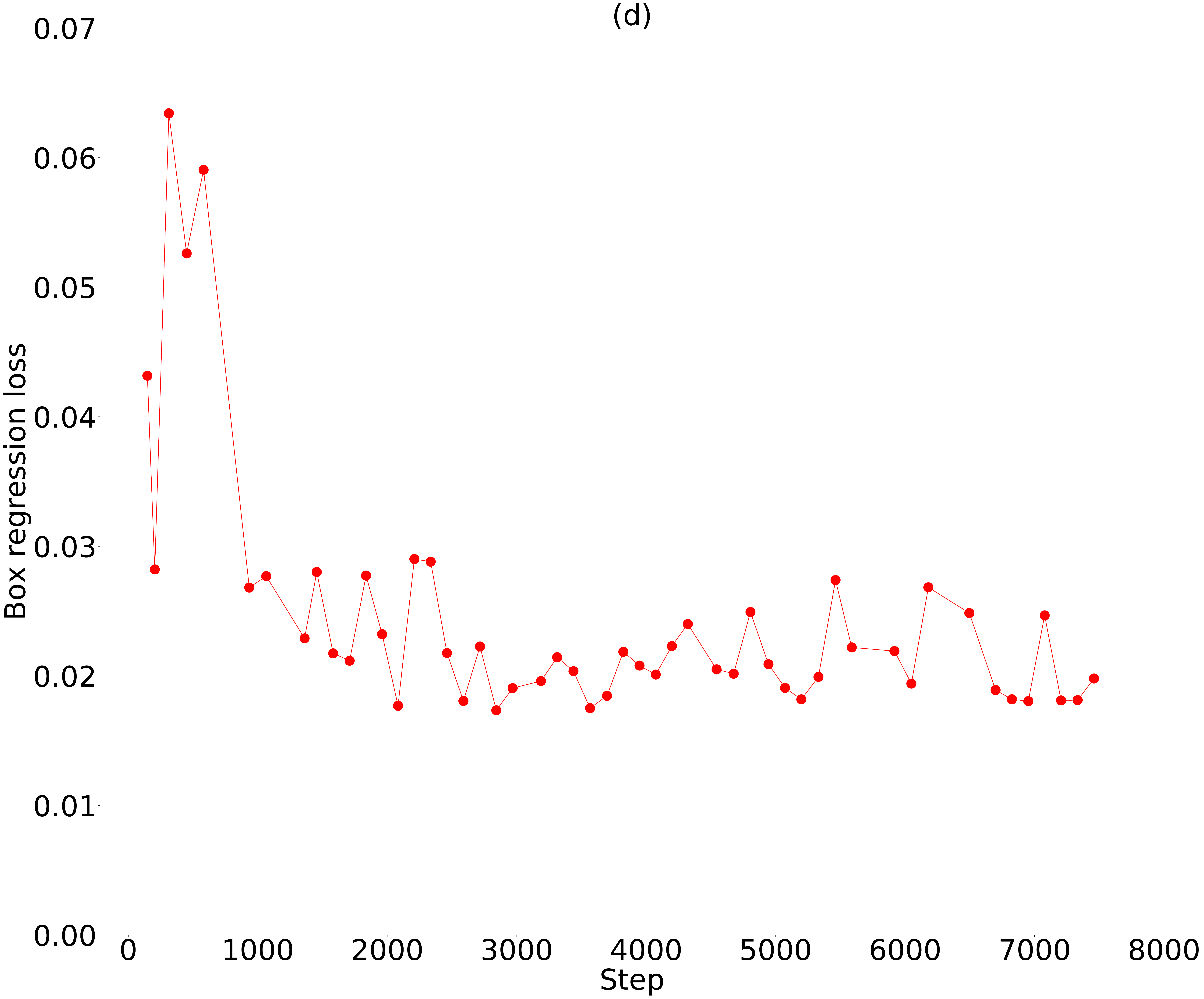}
    \caption{Training losses of the RPN and the DN are represented in (a), (b), (c) and (d). Where (a) displays the RPN objectness loss, (b) displays the RPN box regression loss, (c) displays the DN classification loss and (d) displays the DN box regression loss. The training of the model is stopped after evaluation at 7458 steps when the total loss no longer fluctuates, see Figure \ref{fig:total_loss}. The values for each point in (a),(b),(c) and (d) can be found in Table \ref{tab:all_losses_table}.}
    \label{fig:all_loss}
\end{figure*}

We monitor the performance of the RPN and the DN via their respective loss functions, where we measure the objectness and box regression loss in the RPN and the classification and box regression loss in the DN. Objectness loss measures whether a box is likely to contain a ground truth object, box regression loss measures the exactness of the dimensions between the positive labelled boxes and ground truth box, and classification loss compares the resemblance of the features in a box with the features of the ground truth box. A lower loss value means that the prediction is almost identical to the ground truth. In Figure \ref{fig:all_loss}, we find that each of the losses also eventually stabilise at approximately 3,000 steps. Figure \ref{fig:all_loss} suggests that the RPN is better generalised than the DN as there are fewer fluctuations. This could be explained by two possible reasons: either the training set is suited for localising objects in an image but improvements could be made to enhance the feature classification of objects in the training set, or we need to allow our model to train for more steps. 

The 204 Abell galaxy clusters in the test set appear in the training set at least once but since the positions of the galaxy clusters and their surrounding image environments are different, we can assume the images to be unique. This is a useful test of the localisation performance of our model.

To evaluate our model we use common metrics such as precision, recall and F1-score \citep{metrics}. We do not account for true negatives, as there may be many other astronomical objects aside from galaxy clusters to consider. We expect that finding a galaxy cluster would be relatively rare as stars, galaxies and image artifacts populate the field-of-view. Our approach only searches for clusters rather than attempting to classify all objects. The final output of our model is a `confidence' score generated for every predicted box, where a high confidence score means a high probability of an object being a `real' galaxy cluster. We want to determine a threshold for the confidence score that returns high precision and high recall ratios. We re-run our model on the test set using different confidence score thresholds to examine the number of true positives (TP), false positives (FP) and false negatives (FN) returned.

We define a distance threshold by calculating a linear distance between the predicted and ground truth centre coordinates, where the predicted cluster centre is assumed to be at the same redshift as the ground truth cluster centre. We only apply this distance threshold during the model analysis, as we want to distinguish whether a predicted object is considered as a TP or FP detection to assess classification and localisation. 

For our model, TP refers to the number of predicted boxes that score greater than the confidence score threshold and has a predicted centre within the distance threshold of the ground truth centre. FP refers to the number of predicted boxes that score greater than the confidence score threshold but does not have a predicted centre within the distance threshold of the ground truth centre. FN refers to the number of ground truth galaxy clusters without a positive prediction above the confidence score threshold within the distance threshold.

We calculate the precision and recall ratios using the number of TP, FP and FN at each confidence score threshold. Precision (also known as purity) is a ratio that effectively determines the number of ground truth objects returned by our model compared with the number of new predictions, see Equation \ref{eq:precision}:

\begin{equation}
    \text{Precision} = \frac{\text{TP}}{\text{TP} + \text{FP}}.
	\label{eq:precision}
\end{equation}

Recall (also known as completeness) determines the number of ground truth objects returned by our model but compared with the number of ground truth objects that our model failed to predict, see Equation \ref{eq:recall}:
 
\begin{equation}
    \text{Recall} = \frac{\text{TP}}{\text{TP} + \text{FN}}.
	\label{eq:recall}
\end{equation}

Precision-Recall (PR) curves are used as visual representations to examine the performance of a model, especially when a class population imbalance exists in the dataset \citep{PR_curve}. Each point on the PR curve refers to the precision and recall ratio at a specific cut-off threshold. We explore eleven cut-off thresholds for confidence scores ranging from 0 to 100 per cent and calculate the corresponding F1-score at each confidence score threshold. 

F1-score is the harmonic mean between the precision and recall ratios at each confidence score threshold \citep{F1_score}. We want to maximise the F1-score for our model to find the optimal balance between precision and recall. F1-score is described in Equation \ref{eq:F1-score}:

\begin{equation}
    \text{F1-score}=2 \ \times \ \frac{\text{Precision} \ \times \ \text{Recall}}{\text{Precision} + \text{Recall}}.
	\label{eq:F1-score}
\end{equation}

\begin{figure*}
	\includegraphics[width=\linewidth]{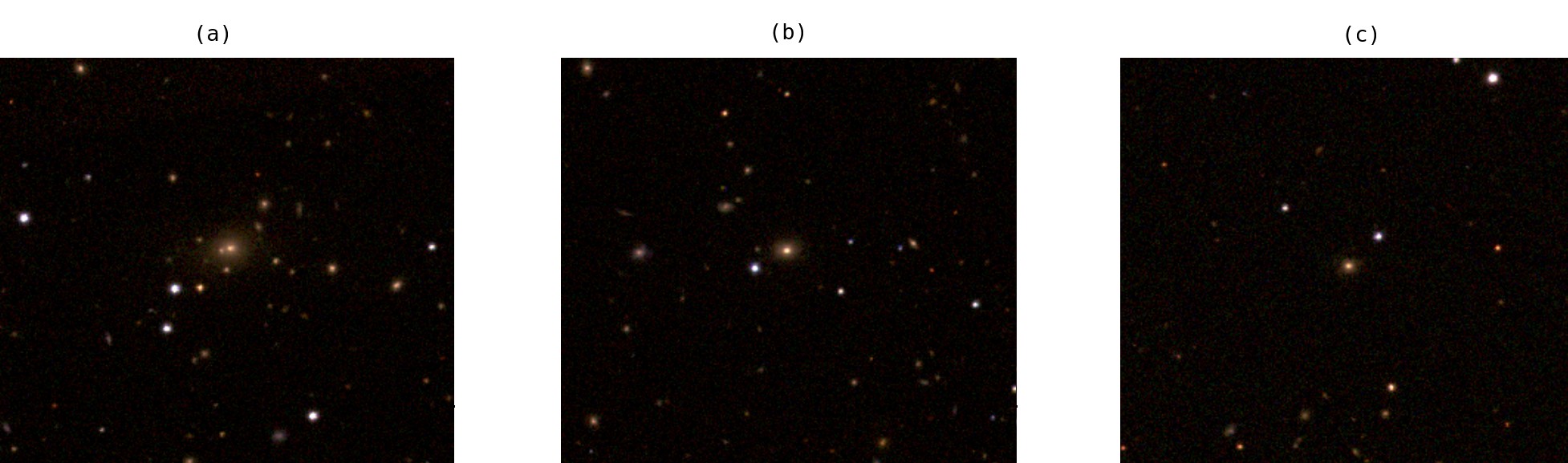}
    \caption{Colour images (a), (b) and (c) contain three different Abell galaxy clusters from the test set. The J2000 coordinates for each galaxy cluster are as follows. (a) RA: 222.78917 and Dec: 14.61203, (b) RA: 180.19902 and Dec: 35.58229 and (c) RA: 137.49464 and Dec: 60.32841. The predicted confidence scores and properties for the galaxy clusters in (a), (b) and (c) can be found in Table \ref{tab:confidence_table}.}
    \label{fig:confidence_score_comparison}
\end{figure*}

We analyse three galaxy clusters (a), (b) and (c) in the test set that have contrasting predicted confidence scores. These galaxy clusters can be seen in Figure \ref{fig:confidence_score_comparison}. From Table \ref{tab:confidence_table}, we find that (c) has the lowest confidence score, whilst (a) has the highest. This is because even though (c) has a high richness value, its higher redshift means it has fainter galaxies. Additionally (b) has a lower richness than (c) but is estimated to have a higher confidence score, again because it is at lower redshift. This demonstrates that galaxy clusters at lower redshift with brighter galaxies will receive higher confidence scores than their fainter counterparts at higher redshift. However, a galaxy cluster would also need high richness to achieve a very high confidence score. As a demonstration of our model, we choose a confidence score threshold of 80 per cent for the remainder of this paper.

\begin{table*}
	\caption{The predicted confidence scores and properties of each galaxy cluster in Figure \ref{fig:confidence_score_comparison}.} 
	\begin{tabular}{lrrrr}
		\hline
	    ID & Confidence Score (\%) & Photometric Redshift & $r$-band Magnitude of the BCG & Richness \\
		\hline
		(a) & 98 & 0.1474 & 14.99 & 78.49\\
		(b) & 51 & 0.1303 & 16.19 & 35.90 \\
	    (c) & 20 & 0.1875 & 16.73 & 58.44 \\
		\hline
	\end{tabular}
	\label{tab:confidence_table}
\end{table*}

\begin{figure*}
    \includegraphics[width=0.49\linewidth]{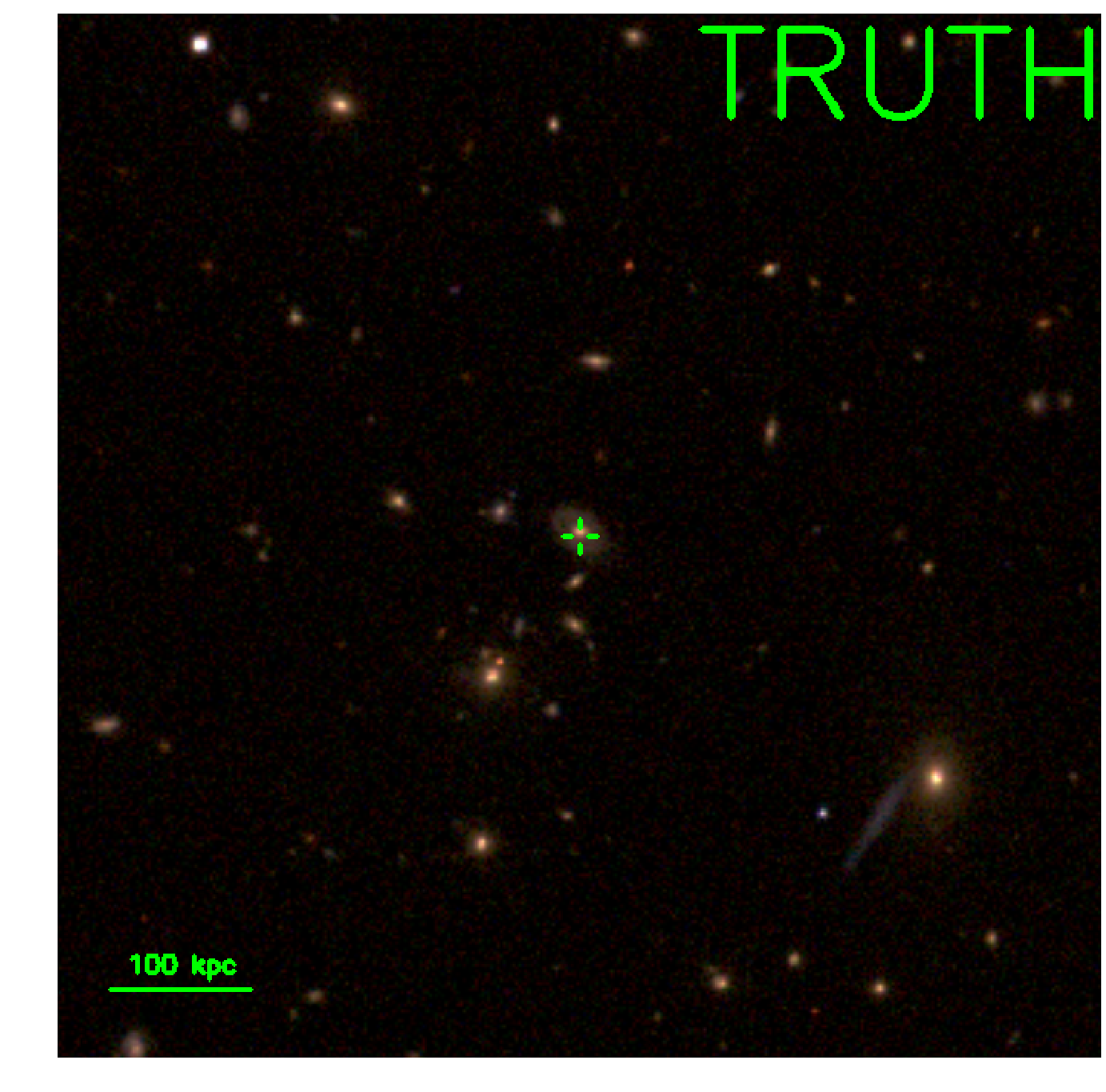}
    \includegraphics[width=0.49\linewidth]{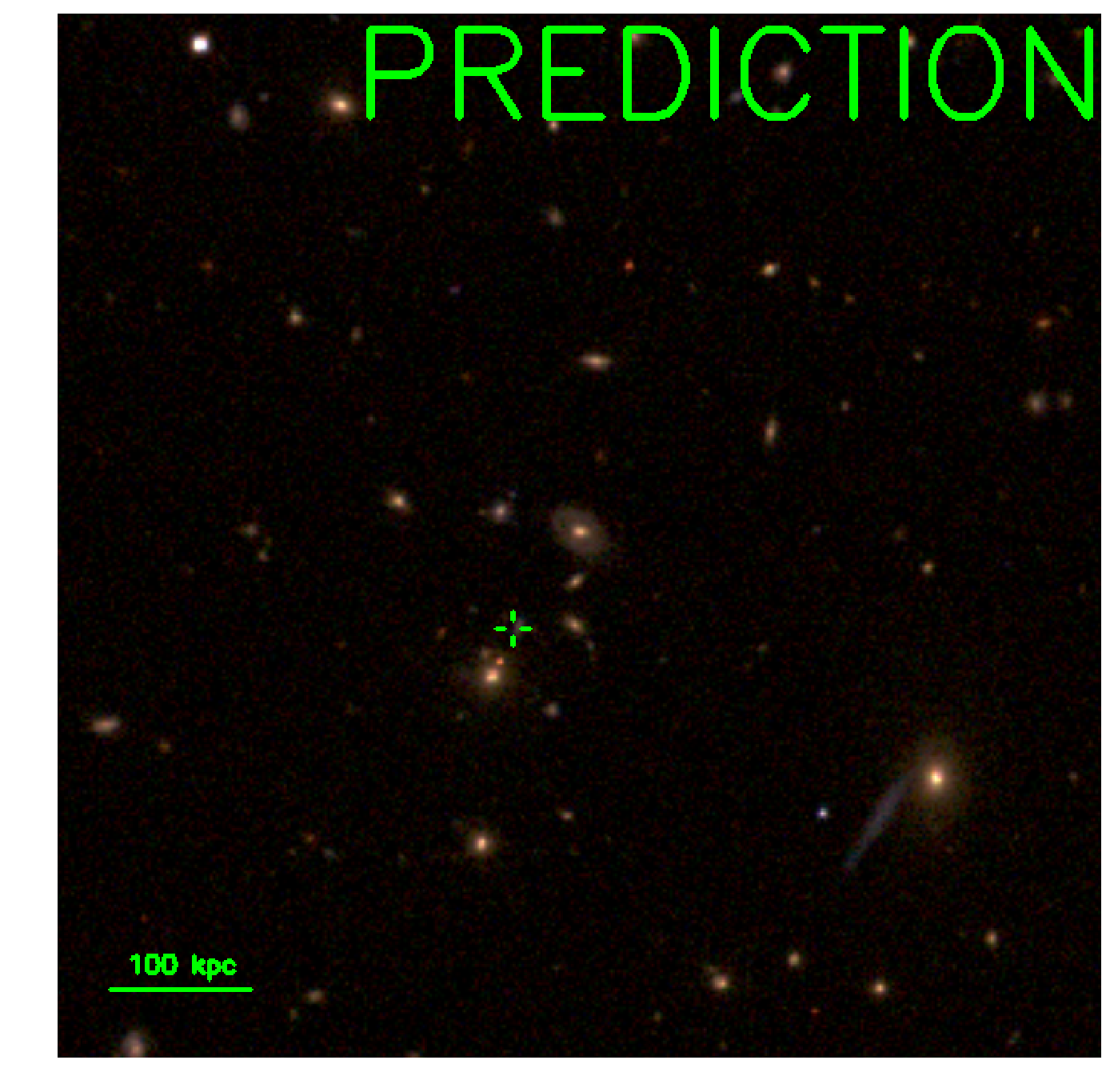}
    \caption{The linear distance between the ground truth and predicted centre coordinate is 88 kpc in respect to the photometric redshift $z=0.1788$ of the ground truth galaxy cluster. The J2000 coordinates of the ground truth galaxy cluster is RA: 191.85623 and Dec: 35.54509.} 
    \label{fig:Test_prediction_distance_small}
\end{figure*}

We investigate how the environment (actual or contaminants) surrounding a galaxy cluster in an image can affect the detections generated by our model. We visually inspect multiple high-scoring candidate galaxy clusters from four different images in the test set. In Figure \ref{fig:Test_prediction_distance_small}, we examine a candidate galaxy cluster that lies just outside the distance threshold at 88 kpc from the ground truth centre. We observe that multiple possible candidate galaxies could be classified as the BCG of the galaxy cluster. Since we train our model to predict a BCG as the galaxy cluster centre we would expect one of the galaxies in Figure \ref{fig:Test_prediction_distance_small} to be chosen. However, we find that our model is unable to definitively determine the ground truth cluster centre if there are multiple BCG-like galaxies close together in an image, such as in the event of an on-going cluster merger. Instead it finds an average centre, which is likely more appropriate for such systems. \cite{relaxation_parameter} developed an approach to determine the dynamical state of a cluster based on photometric data to quantify a relaxation parameter `$\Gamma$', which considers factors such as morphological asymmetry, ridge flatness and normalised deviation. They define the relaxation parameter as $\Gamma \ge 0$ representing as a relaxed state and $\Gamma  < 0$ as a unrelaxed state, such that a larger positive/negative value implies a more relaxed/unrelaxed dynamical state. \cite{relaxation_parameter} suggests that the cluster seen in Figure \ref{fig:Test_prediction_distance_small} is in a very unrelaxed dynamical state with a relaxation parameter value of $\Gamma = -1.43 \pm 0.08$, indicating a possible on-going merger. 

\begin{figure*}
    \includegraphics[width=0.49\linewidth]{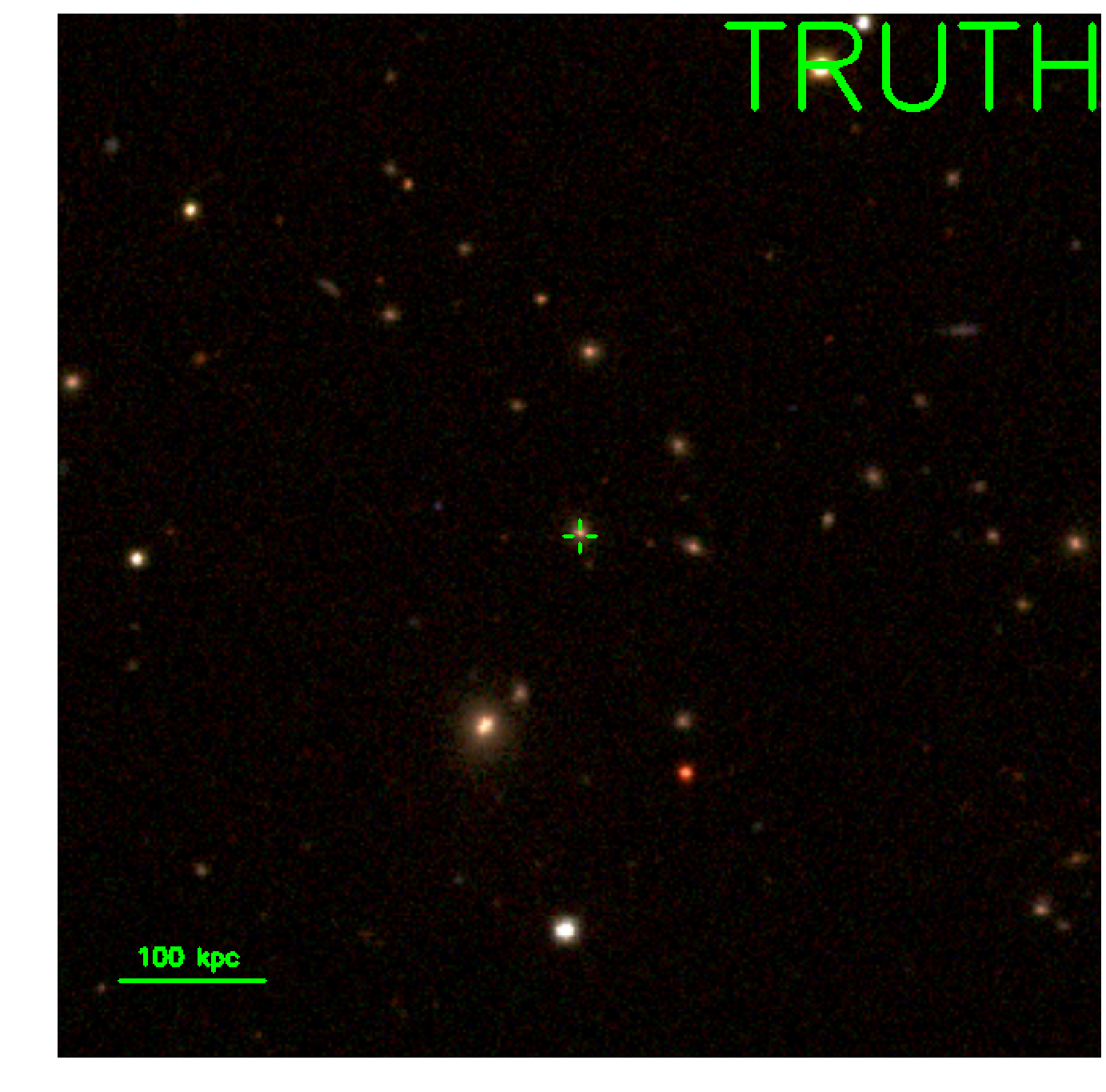}
    \includegraphics[width=0.49\linewidth]{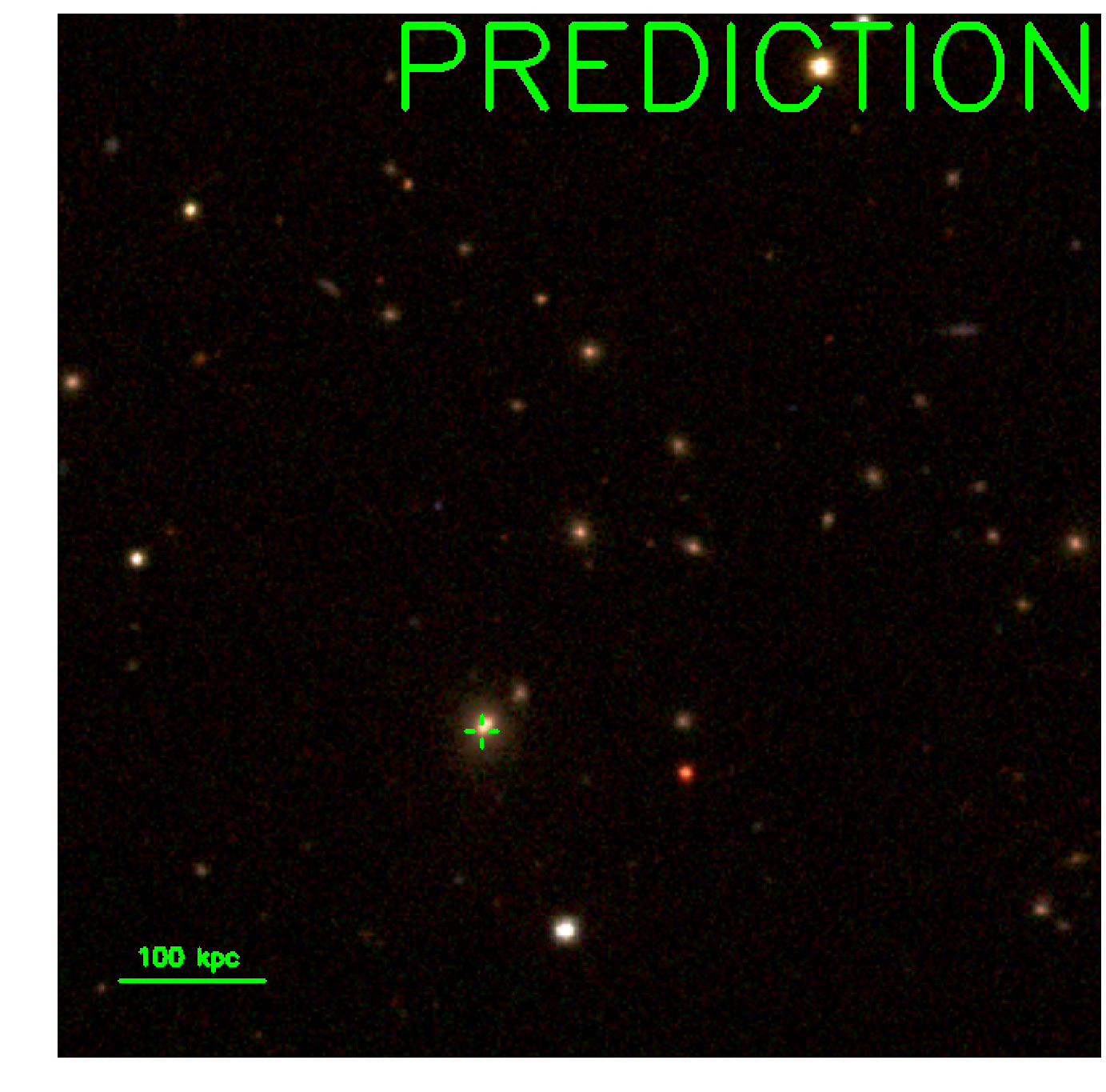}
    \caption{The linear distance between the ground truth and predicted centre coordinate is 158 kpc in respect to the photometric redshift $z=0.1618$ of the ground truth galaxy cluster. The J2000 coordinates of the ground truth galaxy cluster is RA: 161.20657 and Dec: 35.54042.}
    \label{fig:Test_prediction_distance_smallmed}
\end{figure*}

In Figure \ref{fig:Test_prediction_distance_smallmed}, we examine another candidate galaxy cluster that lies 158 kpc from the ground truth centre. We use SDSS-III's Baryon Oscillation Spectroscopic Survey (BOSS, \citealt{sdss_III}) to identify the spectroscopic redshift of the cluster. We identify the spectroscopic redshift of the predicted `BCG' at $z=0.15765$ $\pm$ 0.00003 and the spectroscopic redshift of the ground truth BCG at $z=0.19282$ $\pm$ 0.00003. This means that while the galaxies are in the same line-of-sight they are not part of the same gravitationally-bound system. Our model is unable to determine the ground truth BCG since the predicted `BCG' has stronger visual features and is at a lower redshift. 

\begin{figure*}
    \includegraphics[width=0.49\linewidth]{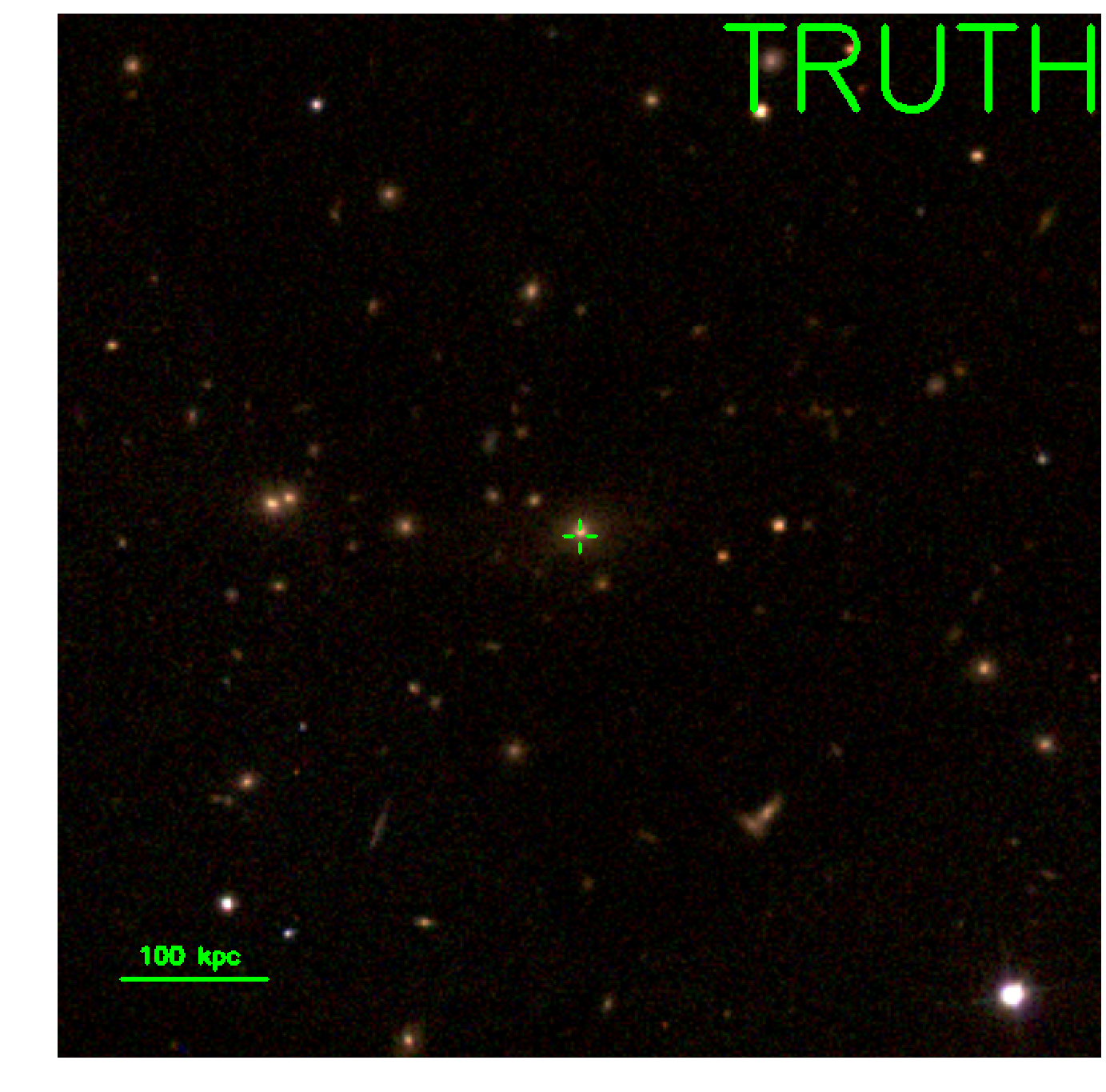}
    \includegraphics[width=0.49\linewidth]{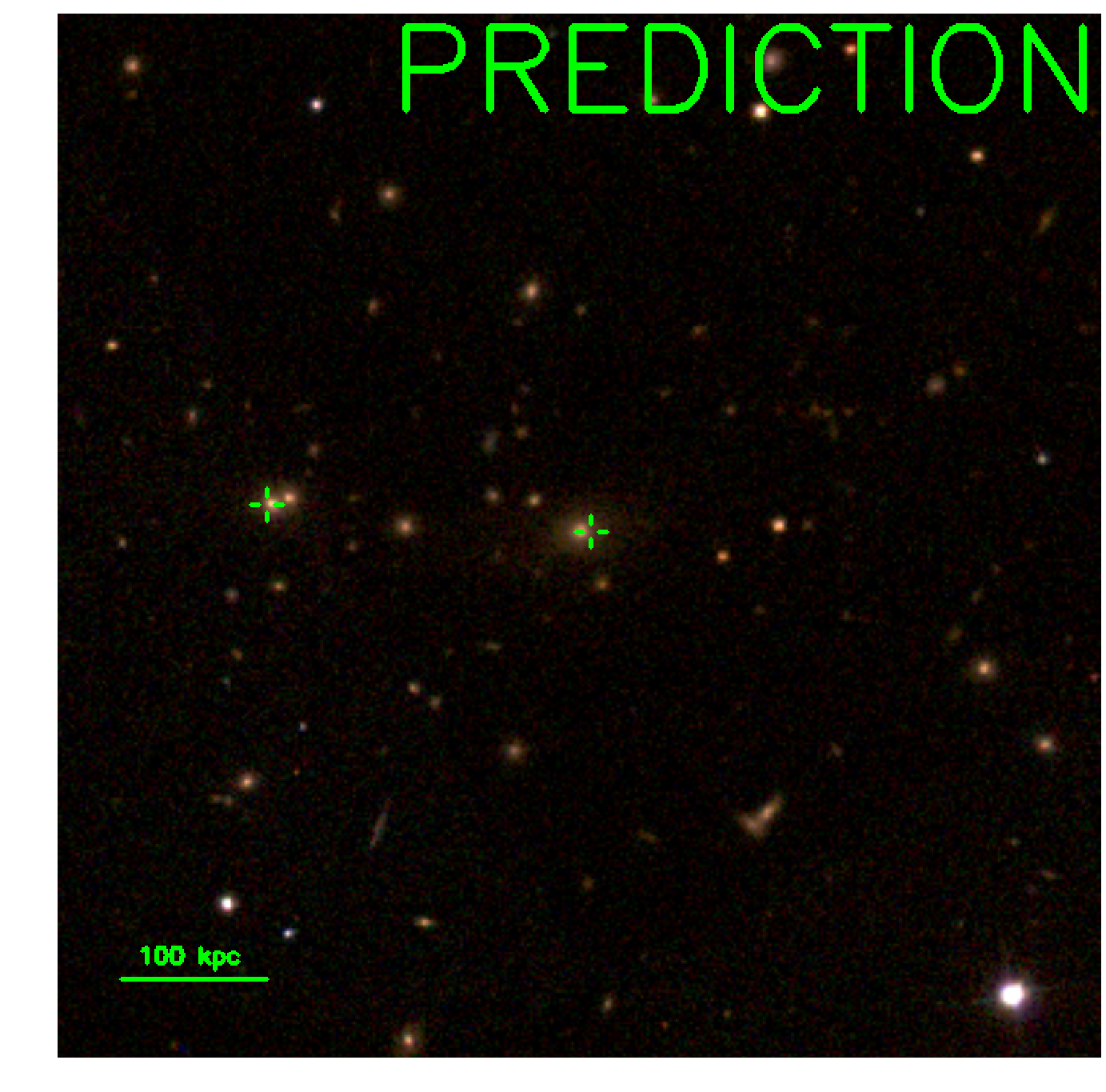}
    \caption{The linear distance between the ground truth and predicted centre coordinate (the one that does not overlap directly with the ground truth) is 220 kpc in respect to the photometric redshift $z=0.1603$ of the ground truth galaxy cluster. The J2000 coordinates of the ground truth galaxy cluster is RA: 353.35867 and Dec: 9.42395.}
    \label{fig:Test_prediction_distance_medium}
\end{figure*}

We find that our model identifies two BCG-like galaxies in Figure \ref{fig:Test_prediction_distance_medium} as potentially two separate galaxy cluster centres but are within each others respective optical mean core radii defined in \S \ref{sec:catalogue_and_image_processing}. One of the two candidate objects lies within the distance threshold of the ground truth centre whilst the other object is 220 kpc away. We find the spectroscopic redshift of the predicted `BCG' is $z=0.15974$ $\pm$ 0.00003 whilst the ground truth BCG has a spectroscopic redshift of $z=0.15989$ $\pm$ 0.00002. \cite{relaxation_parameter} suggests the cluster has a relaxation parameter value of $\Gamma = -0.61 \pm 0.09$. This value implies the cluster is in an unrelaxed dynamical state but to a lesser extent than the cluster seen in Figure \ref{fig:Test_prediction_distance_small}, suggesting that this cluster is not experiencing an on-going merger but is likely in a pre-merger or post-merger state.

\begin{figure*}
    \includegraphics[width=0.49\linewidth]{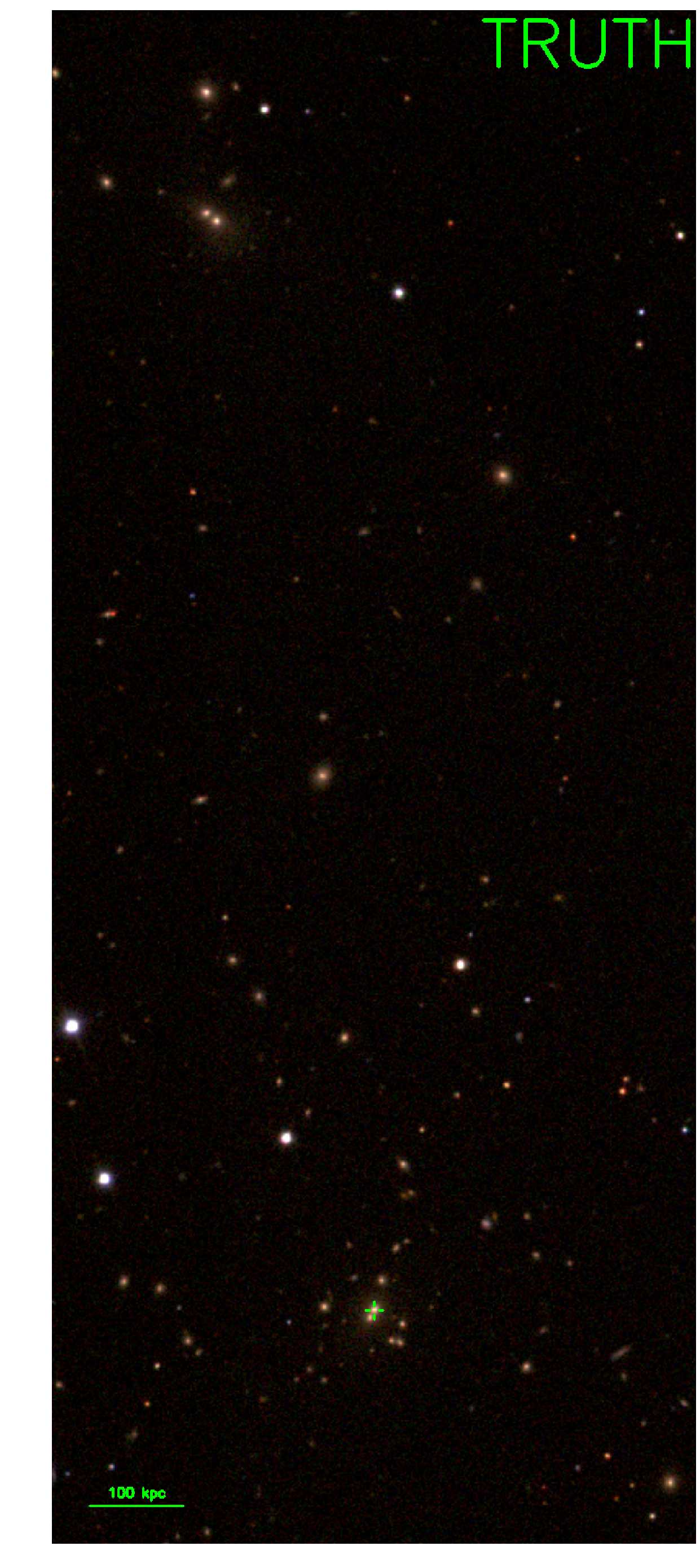}
    \includegraphics[width=0.49\linewidth]{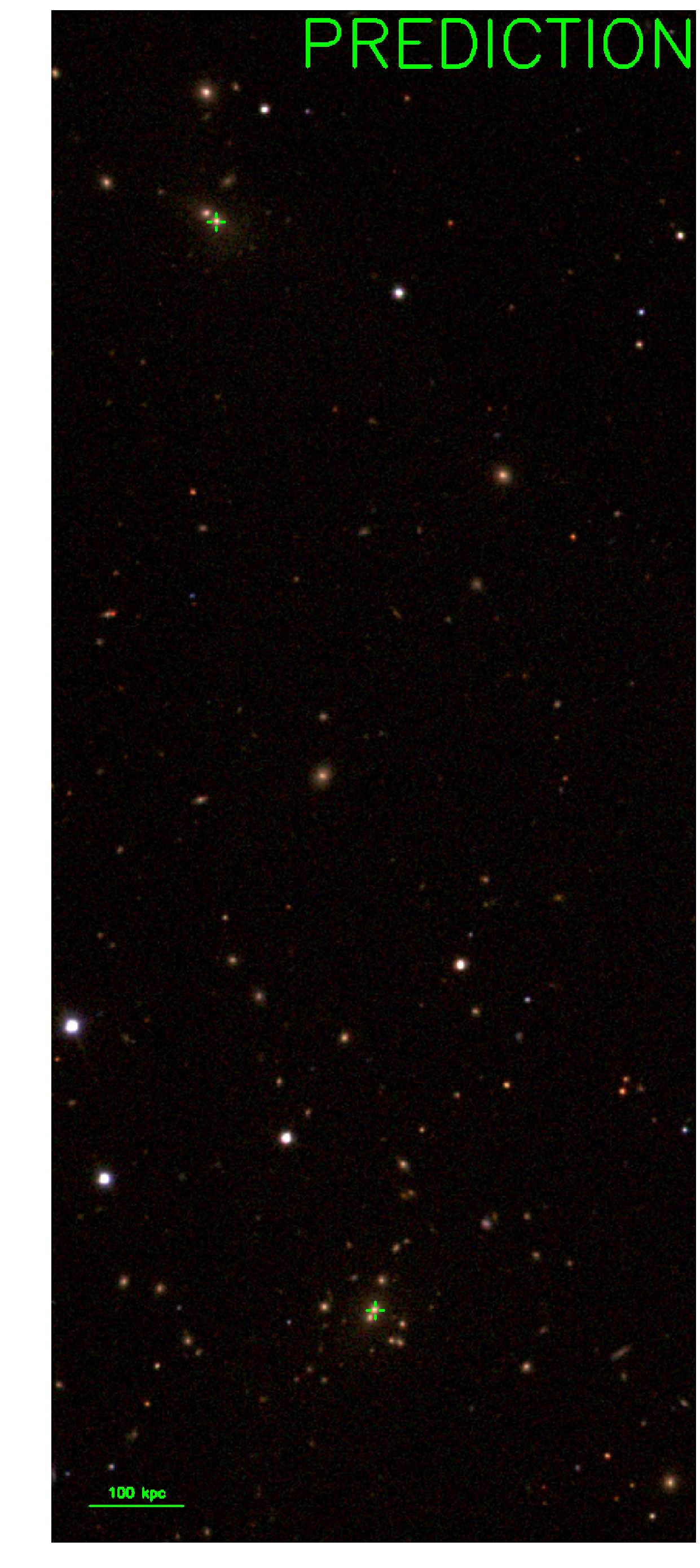}
    \caption{The linear distance between the ground truth and predicted centre coordinate (the one that does not overlap directly with the ground truth) is 1163 kpc in respect to the photometric redshift $z=0.1368$ of the ground truth galaxy cluster. The J2000 coordinates of the ground truth galaxy cluster is RA: 186.96341 and Dec: 63.38483.}
    \label{fig:Test_prediction_distance_large}
\end{figure*}

From Figure \ref{fig:Test_prediction_distance_large}, we find that our model is able to detect a potential galaxy cluster which is extremely far from the ground truth centre at 1163 kpc distance, assuming it is at a similar redshift. We find that the predicted centre is located on top of a BCG-like object. We again use the BOSS survey to find the spectroscopic redshift of the predicted `BCG' at $z=0.10608$ $\pm$ 0.00002 and the ground truth BCG spectroscopic redshift of $z=0.14551$ $\pm$ 0.00003. This shows that the two objects are clearly physically unrelated, therefore we identify the candidate object as a galaxy cluster. We cross-match the new galaxy cluster candidate as `MSPM 08519' with heliocentric redshift of $z=0.1055$ in the \cite{mspm_catalogue} galaxy groups and clusters catalogue.

We decide to set an appropriate distance threshold value based on Figures \ref{fig:Test_prediction_distance_small} and \ref{fig:Test_prediction_distance_smallmed} to be between 88 and 158 kpc. At these distances, we identify multiple galaxy clusters, which are considered far enough apart that cluster mergers and line-of-sight overlap clusters are distinguishable. Since we want to differentiate between cases of TP and FP in our model analysis, we choose a distance threshold of 100 kpc for the remainder of this paper.

In Figures \ref{fig:Test_prediction_distance_medium} and \ref{fig:Test_prediction_distance_large}, we observe secondary positive detections made by our model. We categorise these as FP detections, since the detections do not meet our set threshold criteria. However, we know that FP detections could still be actual galaxy clusters, such as is the case for Figure \ref{fig:Test_prediction_distance_large}. This implies that FP detections consist of candidate galaxy clusters that need further verification or require cross-matching to existing galaxy cluster catalogues. We would label these detections as new galaxy cluster candidates. Whilst in Figure \ref{fig:Test_prediction_distance_smallmed}, we find that no positive detections are made on the ground truth galaxy cluster within the distance threshold. We categorise the ground truth galaxy cluster as a FN and the detection as a FP. This suggests that FN's consist of galaxy clusters that do not appear to have an obvious overdensity of galaxies and do not contain distinctive BCG-like galaxies. These visual features contribute to how distinguishable the galaxy cluster appears at the output of the final layer in the feature network.

\begin{figure}
	\includegraphics[width=\linewidth]{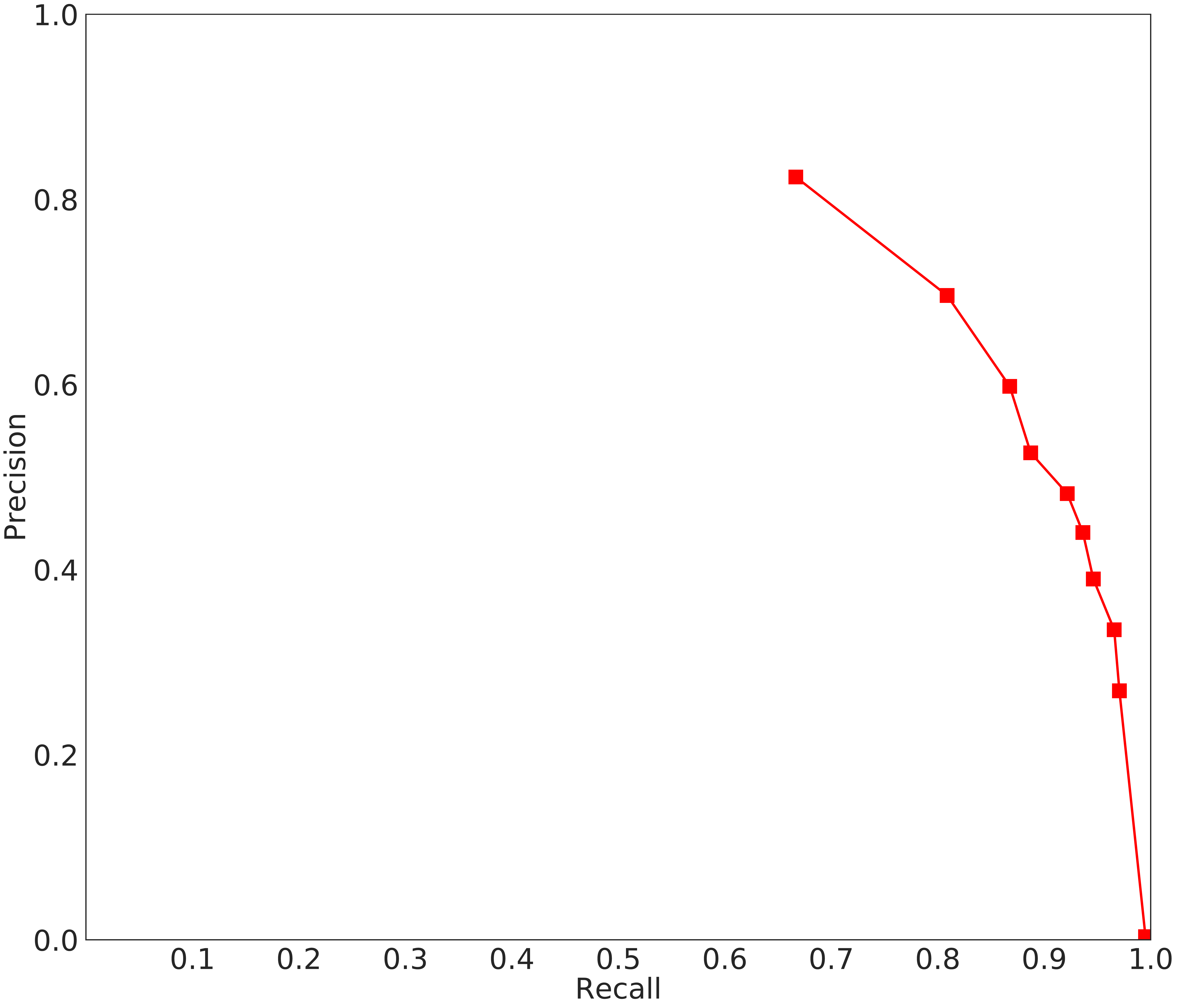}
    \caption{Precision vs Recall ratios from the test set, where each point represents the ratios at a confidence score threshold. The values of each point can be found in Table \ref{tab:confusion_table}. We do not include the precision and recall ratio for the 100 per cent confidence score threshold, as it provides no conclusive evaluation of the performance of the model.}
    \label{fig:PR_curve}
\end{figure}

From Figure \ref{fig:PR_curve}, we observe that high precision diminishes recall and high recall diminishes precision. A low precision ratio results in a large number of candidate objects, whilst a low recall ratio means many real galaxy clusters are not predicted by our model. Table \ref{tab:confusion_table} shows that an 80 per cent confidence score threshold has the highest F1-score, which suggests that this confidence score threshold is the most effective at balancing precision and recall. 

\begin{table*}
	\caption{The total number of true positives, false positives and false negatives returned by our model on the test set, where the precision, recall and F1-score ratios are calculated for each confidence score threshold.} 
	\begin{tabular}{lrrrrrr} 
		\hline
	    Confidence score threshold (\%) & \# TP & \# FP & \# FN & Precision & Recall & F1-score \\
		\hline
		0 & 203 & 60997 & 1 & 0.003317 & 0.9951 & 0.006612 \\
		10 & 198 & 538 & 6 & 0.2690 & 0.9706 & 0.4213 \\
		20 & 197 & 391 & 7 & 0.3350 & 0.9657 & 0.4975 \\
		30 & 193 & 302 & 11 & 0.3899 & 0.9461 & 0.5522 \\
		40 & 191 & 243 & 13 & 0.4401 & 0.9363 & 0.5987 \\
		50 & 188 & 202 & 16 & 0.4821 & 0.9216 & 0.6330 \\
		60 & 181 & 163 & 23 & 0.5262 & 0.8873 & 0.6606 \\
		70 & 177 & 119 & 27 & 0.5980 & 0.8676 & 0.7080 \\
		80 & 165 & 72 & 39 & 0.6962 & 0.8088 & 0.7483 \\
        90 & 136 & 29 & 68 & 0.8242 & 0.6667 & 0.7371 \\
        100 & 0 & 0 & 204 & 0.00 & 0.00 & 0.00 \\
		\hline
	\end{tabular}
	\label{tab:confusion_table}
\end{table*}

\begin{figure}
    \includegraphics[width=\linewidth]{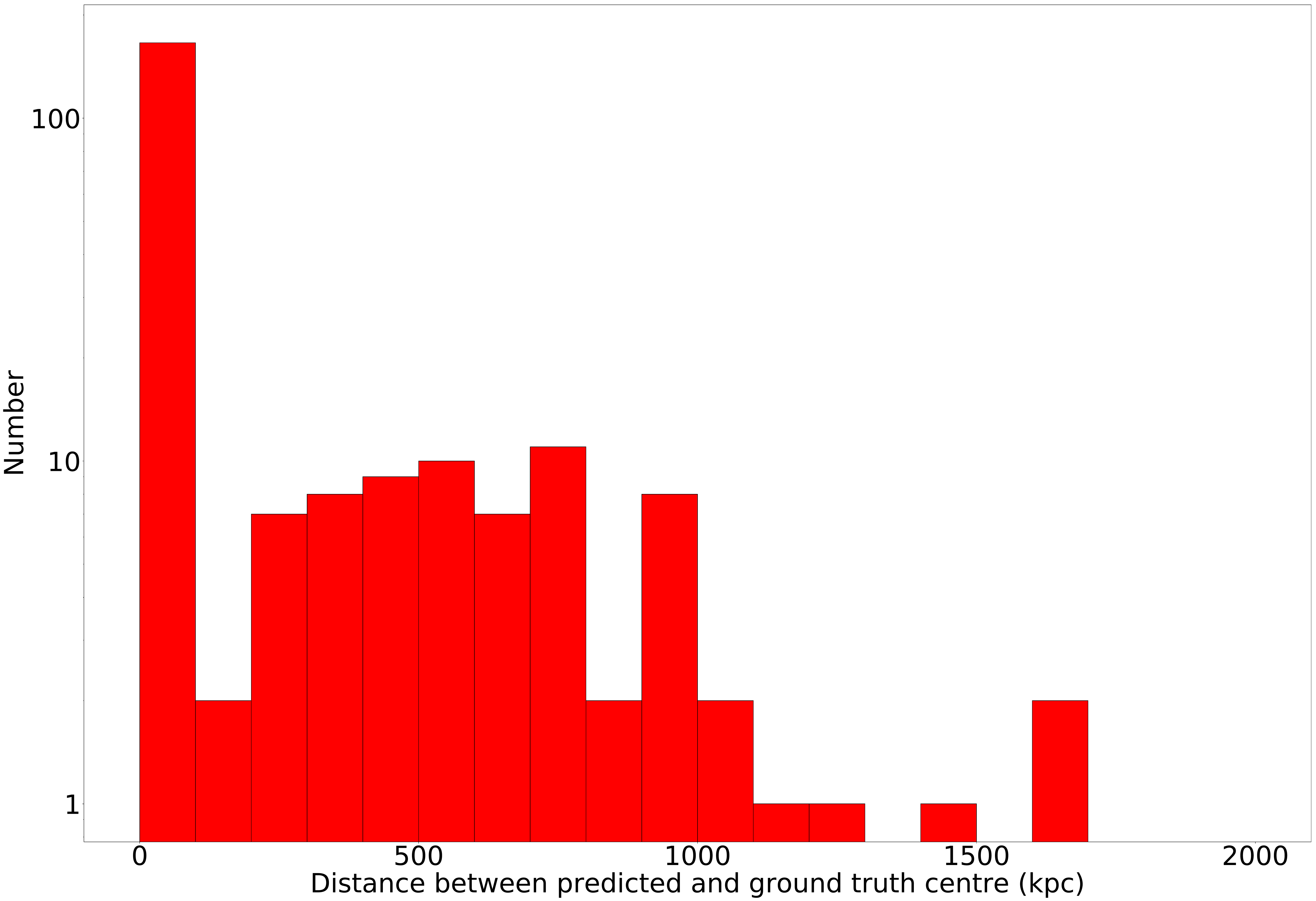}
	\includegraphics[width=\linewidth]{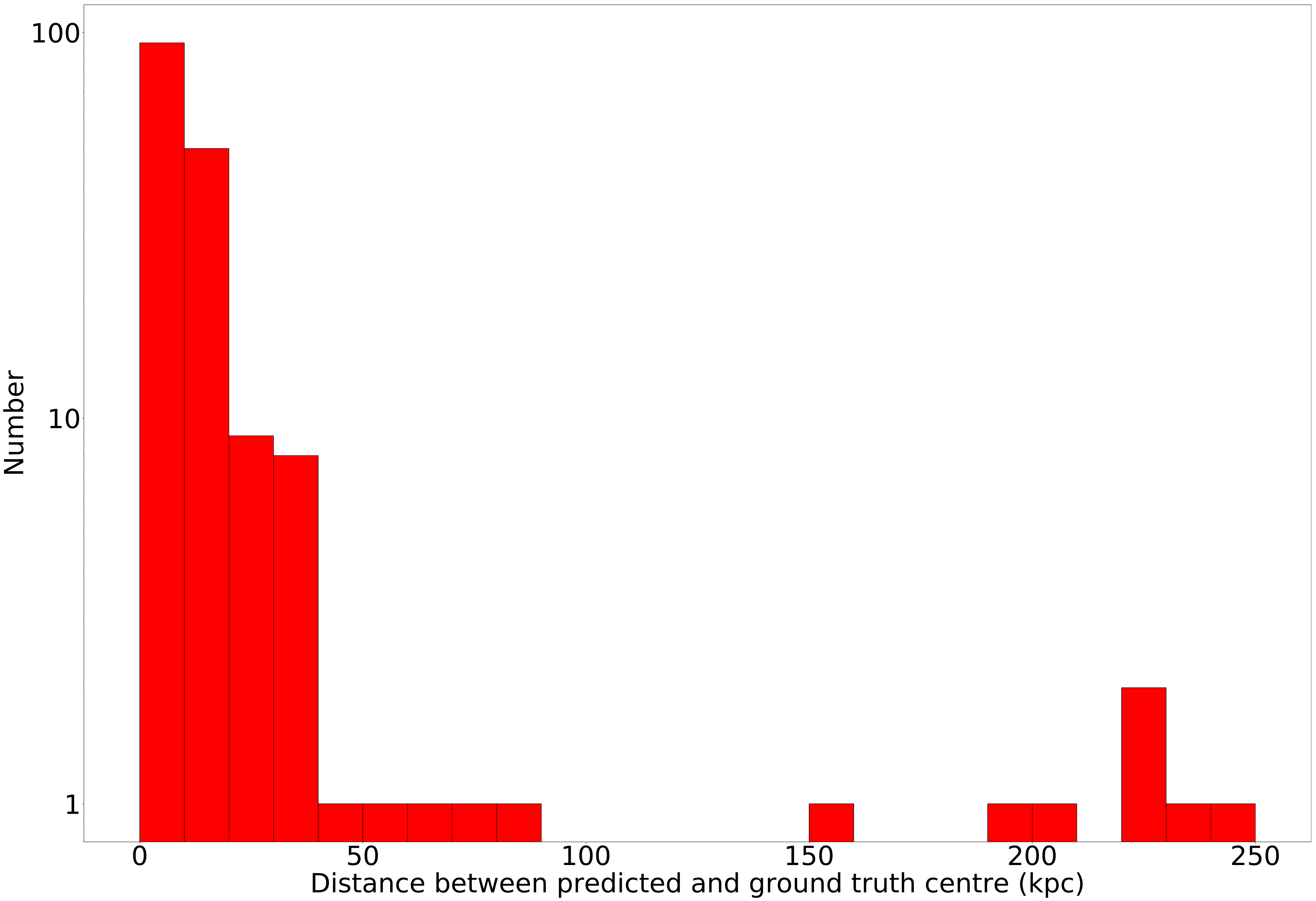}
    \caption{The distribution of the linear distance between the ground truth and predicted centre coordinates in test set images for all predictions (top) and predictions within the distance threshold (bottom) using an 80 per cent confidence score threshold.}
    \label{fig:distance_threshold_plot}
\end{figure}

We analyse the distance between all of the predicted centre coordinates from the ground truth centre using an 80 per cent confidence score threshold. In Figure \ref{fig:distance_threshold_plot}, we find that the distance threshold of 100 kpc contains 70 per cent of all of the predictions and returns 81 per cent of the total ground truth clusters in the test set. We disregard any detection further than 250 kpc from the ground truth centre from being considered as a TP prediction, since the prediction would lie outside the optical mean core radii stated in \S \ref{sec:catalogue_and_image_processing}. We calculate the standard error of coordinate regression of our model to determine the average distance of the predicted centre coordinate from the ground truth centre coordinate using Equation \ref{eq:Standard_error_of_estimate}:

\begin{equation}
    {\sigma}_{estimate} = \sqrt{\frac{\sum{(Y-Y^{'})^{2}}}{N}}, 
	\label{eq:Standard_error_of_estimate}
\end{equation}

where ${\sigma}_{estimate}$ is the standard error of regression, Y is the ground truth value, $Y^{'}$ is the predicted value and N is the sample size \citep{standard_error}. We obtain a standard error of 17.42 kpc for predictions considered as TP's in the test set using an 80 per cent confidence score threshold. Therefore we can estimate a 95 per cent confidence interval for all predicted centre coordinates to be approximately within $\pm1.96 \ \times$ standard error of a ground truth centre coordinate \citep{confidence_interval}. 

\begin{figure}
	\includegraphics[width=\linewidth]{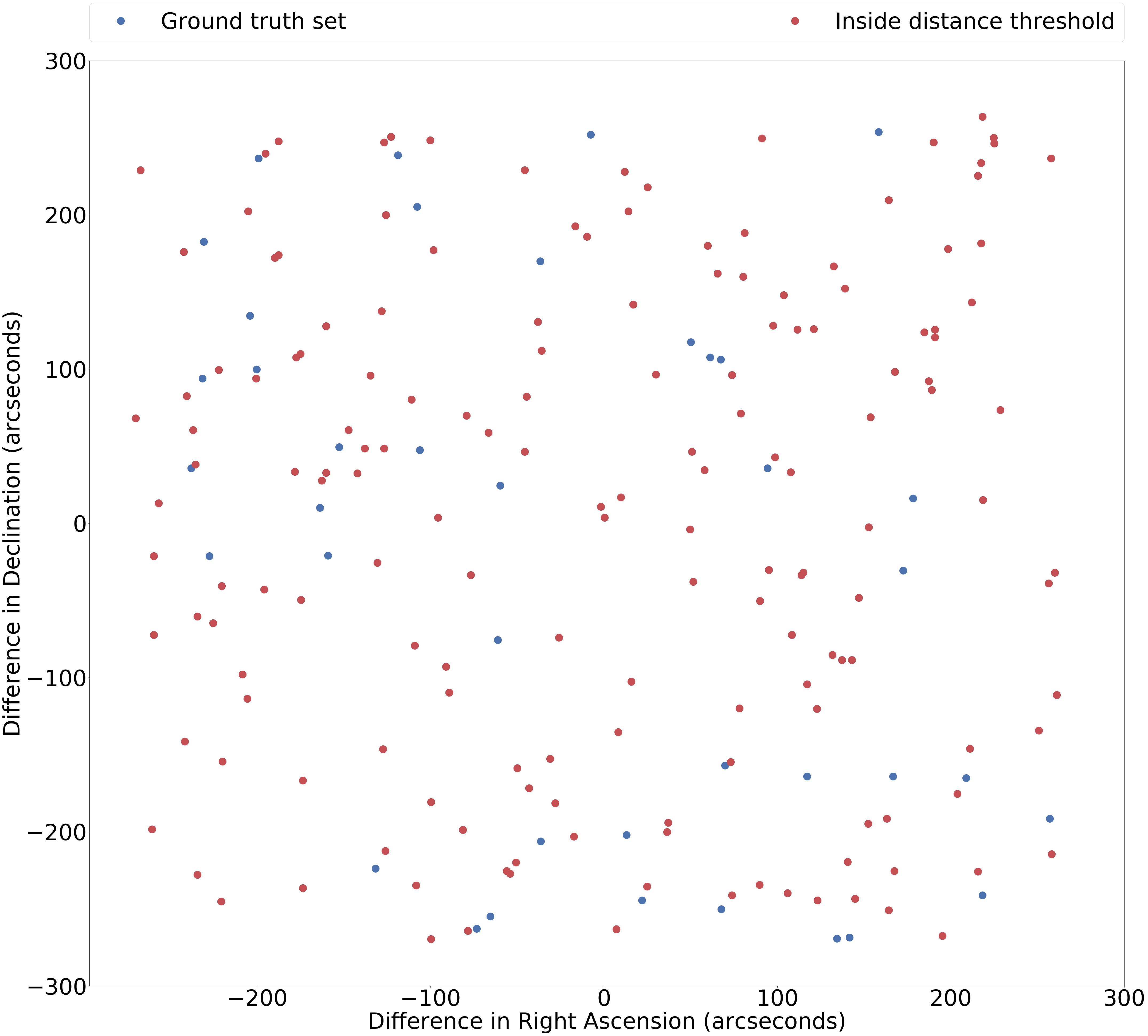}
    \caption{A comparison of the centre coordinate offset between the ground truth galaxy clusters returned by our model and the full list of offset values in the test set (see Figure \ref{fig:image_offset}) using an 80 per cent confidence score threshold.}
    \label{fig:predicted_positions}
\end{figure}

In Figure \ref{fig:predicted_positions}, we visually examine the positions of returned ground truth galaxy clusters in the test set with their original locations as shown in Figure \ref{fig:image_offset}. We find that our model does not show bias towards any particular location in an image. This suggests that the random offset we apply in \S \ref{sec:catalogue_and_image_processing} is effective at reducing location bias during training.

\begin{figure}
	\includegraphics[width=\linewidth]{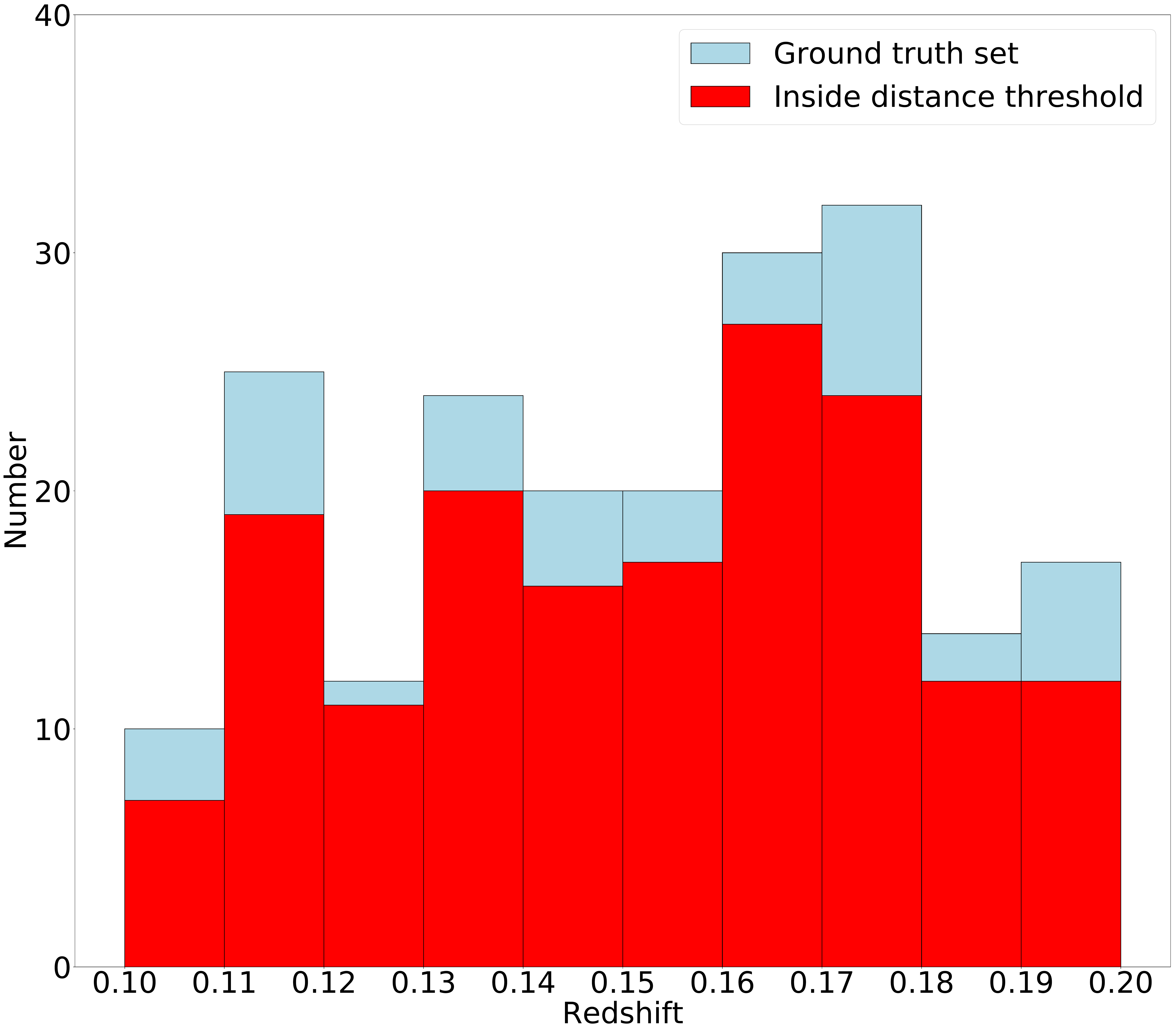}
	\includegraphics[width=\linewidth]{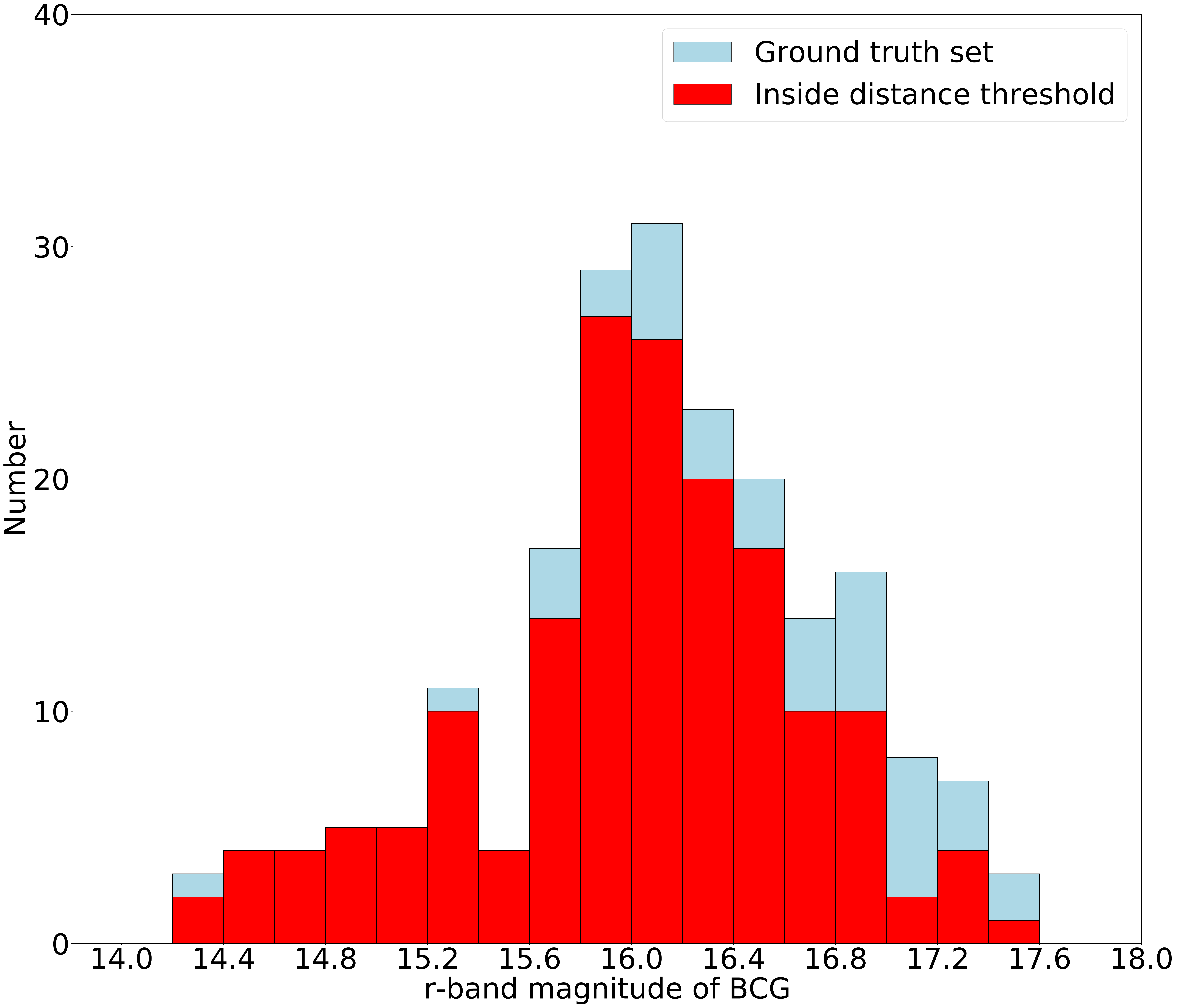}
	\includegraphics[width=\linewidth]{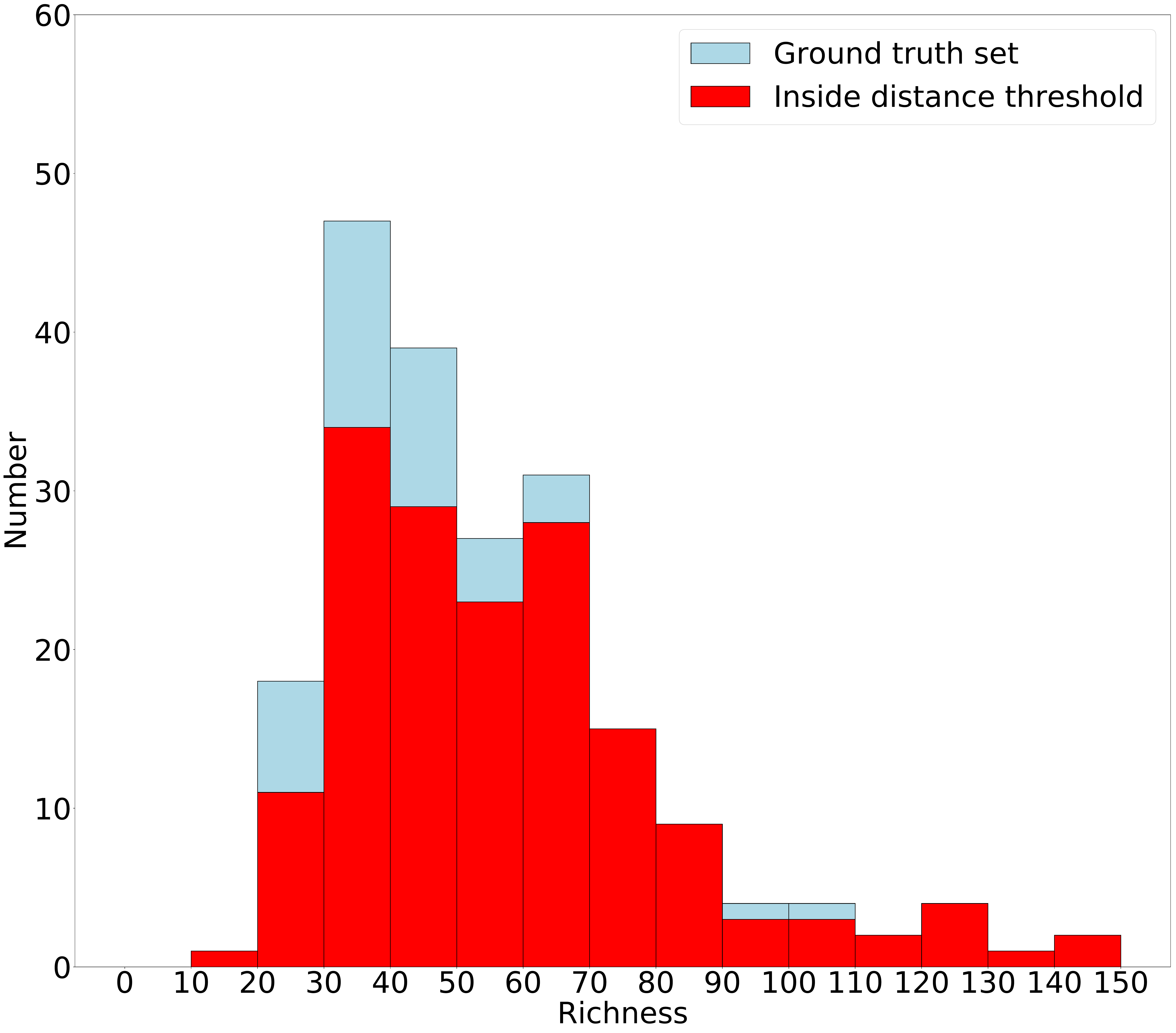}
    \caption{The distributions of the properties from the original and ground truth galaxy clusters returned by our model in the test set using an 80 per cent confidence score threshold. The histograms display the photometric redshift, $r$-band magnitude of the BCG and richness (from top to bottom).}
    \label{fig:test_properties}
\end{figure}

We also compare the photometric redshift, BCG $r$-band magnitude and richness distributions of all the galaxy clusters returned by our model with their original distributions in the test set. Figure \ref{fig:test_properties} shows that our model has no clear prediction bias towards any of these properties. We perform a two sample Kolmogorov-Smirnov (KS) test \citep{ks_test} to test whether the original and returned distributions violate the null hypothesis. Since the KS test is non-parametric, the distributions do not need to have normality. We calculate test statistic values of 0.06275, 0.07335 and 0.02193 for photometric redshift, BCG $r$-band magnitude and richness respectively. We set $\alpha = 0.05$ as the level of significance to obtain a critical value of 0.1281 \citep{ks_test_critical_value}. Since the test statistic values are smaller than the critical value at $\alpha = 0.05$, we cannot reject that the original and returned distributions are statistically the same.

\subsection{Comparison To redMapper galaxy clusters}
\label{Comparison_To_redMapper_galaxy_clusters}

\begin{figure}
	\includegraphics[width=\linewidth]{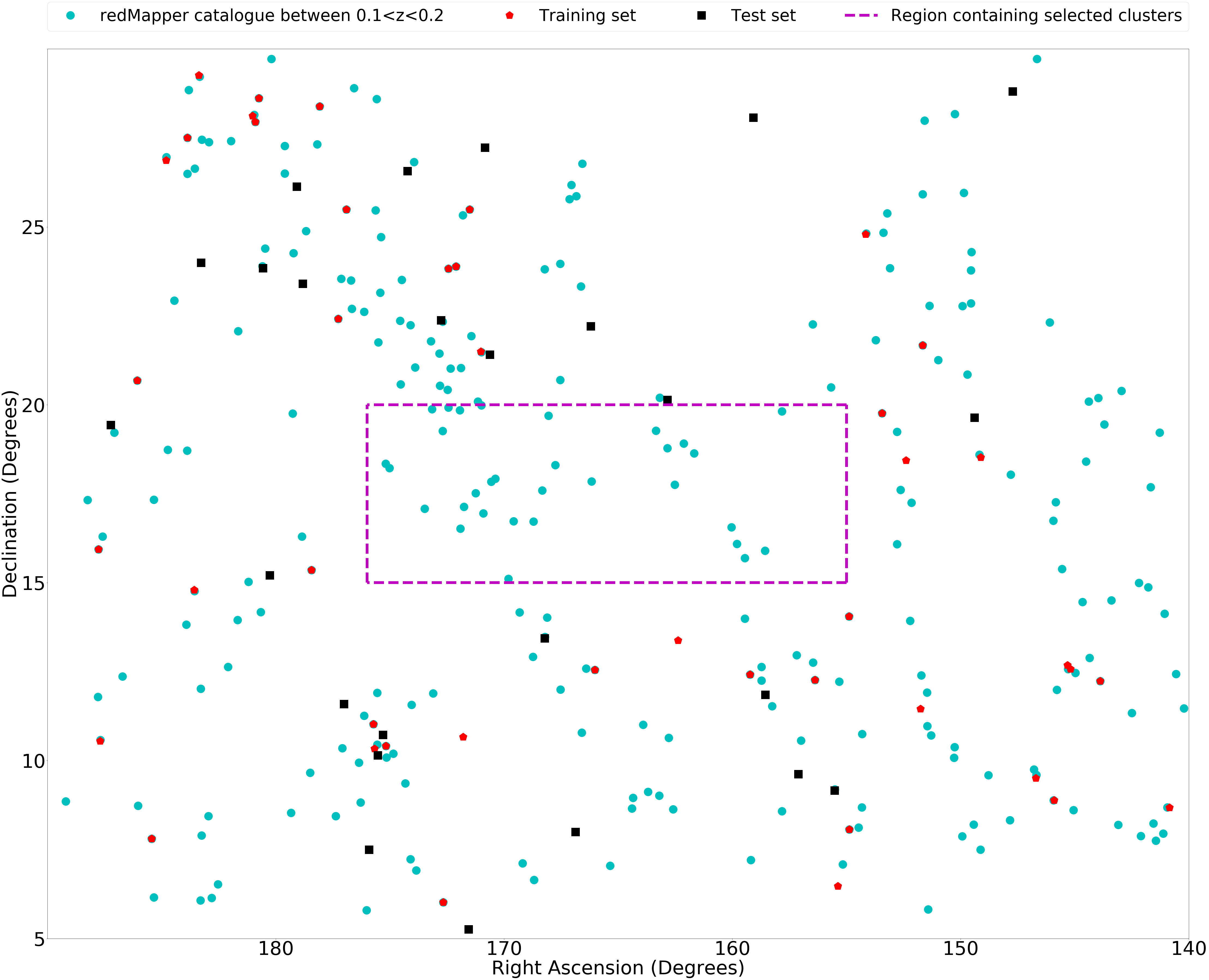}
    \caption{A map of astronomical coordinates using the J2000 epoch system for the galaxy clusters in the training set, test set and constrained redMapper catalogue \protect\citep{redMapper}. We highlight the region of galaxy clusters identified by the redMapper algorithm, which are not already part of the training set or test set.}
    \label{fig:redMapper_region_of_interest}
\end{figure}

The redMapper algorithm predicts galaxy clusters using the red sequence fitting technique and probabilistic percolation of galaxies based on their photometric redshift. \cite{redMapper} apply their algorithm to SDSS DR8 \citep{sdss_dr8}, to create a catalogue of ${\sim}25,000$ candidate galaxy clusters in the photometric redshift range of $0.08 < z < 0.55$. We apply the same testing constraints used in \S \ref{sec:catalogue_and_image_processing} on the redMapper galaxy clusters, where galaxy clusters must be in the photometric redshift range of $0.1 < z < 0.2$. We do not need to apply a galaxy member count constraint as the redMapper algorithm by default only recognises galaxy clusters with greater than 20 member galaxies\footnote[11]{Note that \cite{redMapper} does not define galaxy members within $\text{R}_{200}$ but from an optical radius cutoff that scales with the number of galaxies found via percolation.}. In Figure \ref{fig:redMapper_region_of_interest}, we locate a 105 square degree region that contains 31 galaxy clusters identified by the redMapper algorithm. We found that this region did not contain any galaxy clusters from our training or test sets in the photometric redshift range of $0.1 < z < 0.2$ but it did contain clusters found via redMapper in the same redshift range. We use these redMapper clusters to create a new sample test set and examine the localisation and classification performance of our model on unseen clusters.

\begin{figure}
	\includegraphics[width=\linewidth]{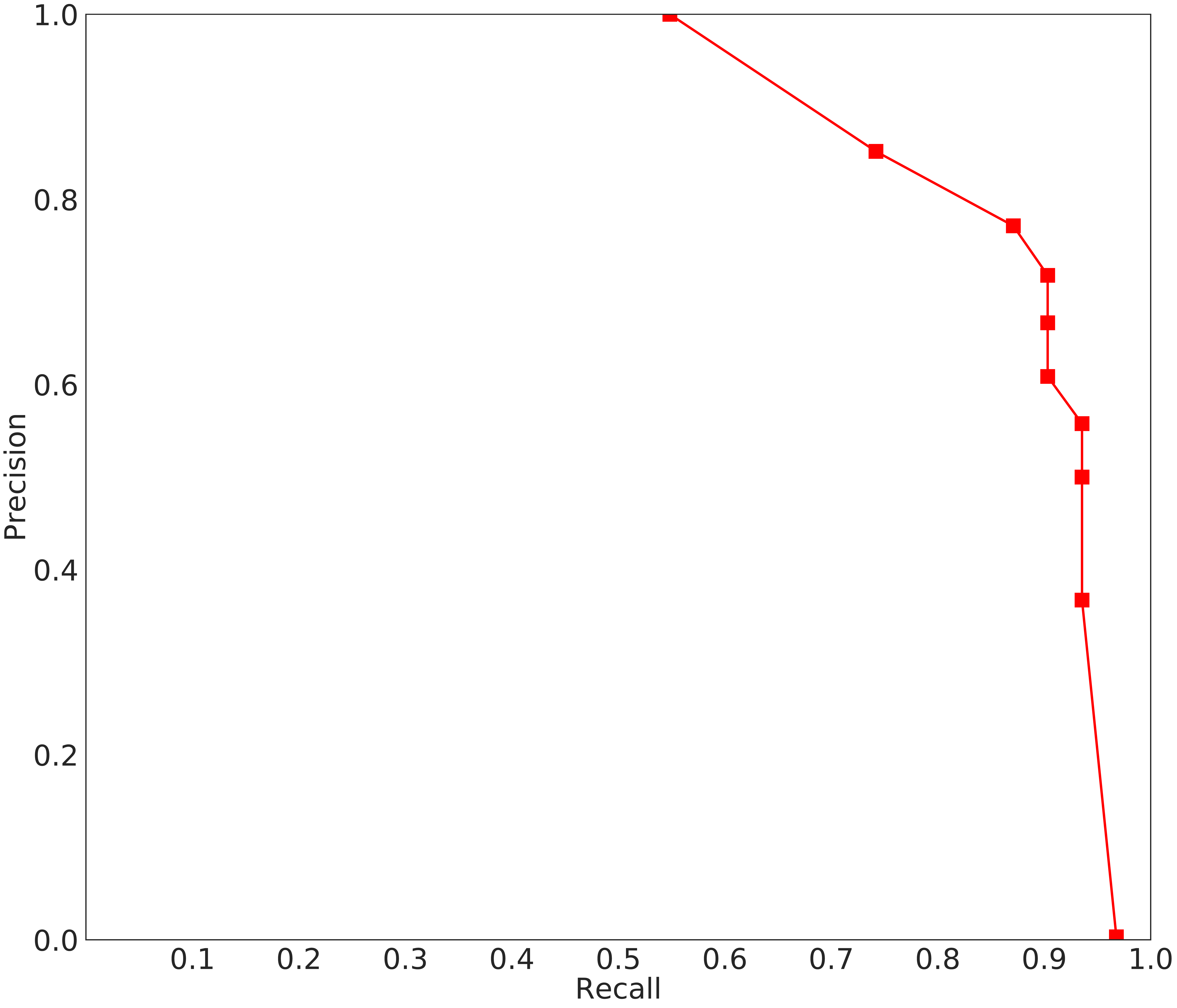}
    \caption{Precision vs Recall ratios on the redMapper test set, where each point represents the ratios at a confidence score threshold. The values of each point can be found in Table \ref{tab:redMapper_confusion_table}. Again we do not include the precision and recall ratio for the 100 per cent confidence score threshold, as it provides no conclusive evaluation of the performance of the model.}
    \label{fig:redMapper_PR_curve}
\end{figure}

We adopt the same procedure from \S \ref{sec:catalogue_and_image_processing} to generate new redMapper test set images. We re-run our model on the redMapper test set and apply the evaluation metrics of precision, recall and F1-score again. In Figure \ref{fig:redMapper_PR_curve}, we observe a precision and recall trade-off similar to Figure \ref{fig:PR_curve}, where precision increases with confidence score threshold whilst recall decreases. From Table \ref{tab:redMapper_confusion_table}, we find the confidence score with the highest F1-score is 70 per cent. This suggests that our model has not overfit since it performs better on an unseen dataset, where the confidence score threshold has lowered from 80 per cent in \S \ref{Model_Analysis_with_Test_Set}. A lower confidence score threshold increases the number of objects detected while still retaining high precision and recall. From using a 70 per cent confidence score threshold, we obtain a standard error of 12.33 kpc for predictions considered as true positives in the redMapper test set.

\begin{table*}
	\caption{The total number of true positives, false positives and false negatives returned by our model on the redMapper test set where the precision, recall and F1-score ratios are calculated for each confidence score threshold.} 
	\begin{tabular}{lrrrrrr} % 8 columns, alignment for each
		\hline
	    Confidence score threshold (\%) & \# TP & \# FP & \# FN & Precision & Recall & F1-score \\
		\hline
		0 & 30 & 9270 & 1 & 0.003226 & 0.9677 & 0.006430 \\
		10 & 29 & 50 & 2 & 0.3671 & 0.9355 & 0.5273 \\
		20 & 29 & 29 & 2 & 0.50 & 0.9355 & 0.6517 \\
		30 & 29 & 23 & 2 & 0.5577 & 0.9355 & 0.6988 \\
		40 & 28 & 18 & 3 & 0.6087 & 0.9032 & 0.7273 \\
		50 & 28 & 14 & 3 & 0.6667 & 0.9032 & 0.7671 \\
		60 & 28 & 11 & 3 & 0.7179 & 0.9032 & 0.80 \\
		70 & 27 & 8 & 4 & 0.7714 & 0.8710 & 0.8182 \\
		80 & 23 & 4 & 8 & 0.8519 & 0.7419 & 0.7931 \\
        90 & 17 & 0 & 14 & 1.00 & 0.5484 & 0.7083 \\
        100 & 0 & 0 & 31 & 0.00 & 0.00 & 0.00 \\
		\hline
	\end{tabular}
	\label{tab:redMapper_confusion_table}
\end{table*}

We follow-up the FP detections made on the redMapper test set images by cross-matching with the redMapper and \cite{clusters_catalogue} catalogues with relaxed redshift constraints. We use the 70 per cent confidence score threshold, as this threshold has the highest F1-score (see Table \ref{tab:redMapper_confusion_table}), from which we obtain eight new candidate galaxy clusters. We apply a redshift constraint of $0.05 < z < 0.4$ on the galaxy clusters in the redMapper and \cite{clusters_catalogue} catalogues. We utilise a ${\sim}1.61$ arcminute radius of the predicted RA and Dec coordinates, which corresponds with the optical mean core radii of 250 kpc for a galaxy cluster at redshift $z=0.15$, to search through the catalogues. We find that none of the detections match with existing clusters in the redMapper catalogue. However, we identify three FP detections as existing galaxy clusters in the \cite{clusters_catalogue} catalogue. This shows that our model is generalised enough to detect galaxy clusters that have not been drawn from the same sample as the training set. We find five detections that do not match with any known galaxy cluster in the redMapper and \cite{clusters_catalogue} catalogues but these may exist in other galaxy cluster catalogues.

\section{Discussion}
\label{Discussion}

\subsection{Limitations of our model}
\label{Limitations_of_our_model}

Feature engineering is an important process for improving computational efficiency and the performance of a deep learning model. We apply constraints to the training set to reduce the complexity of the features. For example, we use Abell galaxy clusters which contain a minimum of 50 galaxies within a $1.5 \ \text{h}^{-1}$ Mpc radii of the galaxy cluster centre \citep{abell_clusters}. This means that these galaxy clusters would have strong signal-to-noise and are very likely to be real gravitationally bound clusters. However, not all Abell galaxy clusters have been verified so one limitation of our approach is that our model is reliant on the catalogue from \cite{clusters_catalogue} for training data. In \cite{clusters_catalogue}, Monte Carlo simulations are used to determine a false detection rate of less than $6$ per cent for the entire catalogue. False detections lower the overall precision of the predictions because the training data could be contaminated with objects that should not be classified as galaxy clusters. Similarly, the test set will also suffer from this. Since we have a large training set, it is impractical to directly check for contaminants in every image with spectroscopic follow-up of all ground truth cluster members. We must therefore assume that all of the galaxy clusters in this catalogue are real. Additionally we must account for the errors in RA and Dec coordinates from \cite{clusters_catalogue}, as we use these coordinates for the ground truth centre coordinates. 

As with all deep learning algorithms, there are hyper-parameters that require either minor or extensive fine tuning, depending on the task at hand. The main training hyper-parameters of Faster-RCNN include the learning rate, momentum, gradient clipping threshold, mini-batch size, number of layers, number of neurons in each layer and architecture. We have shown that the values set for these hyper-parameters are capable of being adapted to perform generalised object detection of galaxy clusters. However, to fully optimise the value of every hyper-parameter is computationally expensive, so we rely on the use of transfer learning for partial optimisation of the hyper-parameters in our deep learning model.

We adopt a specific methodology to generate all of the images by applying the same contrasting, image aspect and image scaling ratios for computational efficiency. However, this may create a trade-off between computational efficiency and bias from image pre-processing. This means that all future input images to our model are somewhat restricted to using the same pre-processing techniques that we apply in \S \ref{sec:catalogue_and_image_processing} to obtain maximum performance.

We perform hold-out validation on the sample set to form the training and test sets. However, this approach is limited to a simple approximation since we observe a population imbalance in the sample in Figure \ref{fig:train_properties}. For example, we find that there are fewer low redshift galaxy clusters compared with high redshift galaxy clusters. This means our model could overfit to populations that appear more frequently. To reduce the chance of overfitting from population bias in the training set we could perform k-fold cross validation \citep{k_fold_cross_validation} when splitting the sample set. K-fold validation splits the sample set into an arbitrary `k' number of folds where one fold becomes the test set whilst the remaining folds are merged to form the training set. This is repeated until every fold has been used as the test set. Then all the folds are compared and the fold with the best performance is chosen to represent the training and test set.

\subsection{Future Applications of this Technique}
\label{future_work}

LSST and \textit{Euclid} are ideal surveys to apply our deep learning model to, as they will be wider and deeper than any survey conducted before them. This will produce detections of many thousands of candidate high redshift or low mass clusters that are currently undiscovered. This may be an iterative process in practice with these large datasets also being used to improve the training of our model.

The Deep-CEE method will be of great use for confirming candidate galaxy clusters detected by X-ray or SZ surveys, as they often have many interlopers. We also intend to use training sets based on galaxy clusters selected from traditional techniques. For example, it will be interesting to compare the galaxy clusters predicted by a deep learning algorithm based on a training set of X-ray selected galaxy clusters compared with those trained on a red sequence fitting sample. This may be a good way to test the various biases of cluster detection methods, which can filter through to any cosmological predictions made with them.

Our deep learning model can also be adapted and applied to both optical imaging and other cluster detection methods at the same time. For example, a related algorithm could be shown a red sequence fit and/or an X-ray image at the same time as visual imaging to create a robust sample of galaxy clusters.  

To improve the model itself, we will investigate whether applying additional image augmentation techniques such as rotation and vertical flipping can improve the performance of our model. \cite{image_augmentation} shows that using simple image transformations can result in a more robust model. We will also train the model to predict intrinsic properties of galaxy clusters such as redshift and richness, which will be vital for the thousands of galaxy clusters discovered by wide-field surveys such as LSST and \textit{Euclid}. This would improve our criteria for categorising TP's.

To prepare for LSST and \textit{Euclid} we aim to run our model on the entirety of SDSS and cross-match all FP detections with existing catalogues. This would be beneficial in constraining the resultant purity ratio of our model, as we expect many FP detections to be actual galaxy clusters.

\section{Conclusion}
\label{Conclusion}

We present Deep-CEE a novel deep learning model for detecting galaxy clusters in colour images and returning their respective RA and Dec. We use Abell galaxy clusters from the \cite{clusters_catalogue} catalogue as the ground truth labels in colour images to create a training set and test set. We determine an optimal confidence score threshold based on the threshold value with the greatest F1-score. We initially find that a 80 per cent confidence score threshold is optimal for finding galaxy clusters in our original test set, as we achieve a precision ratio of 70 per cent and a recall ratio of 81 per cent\footnote[12]{An ideal confidence score threshold would have a precision and recall ratio of 100 per cent. \citep{random_classifier}}. We find that a 70 per cent confidence score threshold is the optimal threshold for detecting galaxy clusters in the redMapper test set, as we obtain a precision ratio of 77 per cent and a recall ratio of 87 per cent. We consider any detections with a predicted confidence score greater than the confidence score threshold as galaxy cluster candidates. It should be noted that our precision ratios are specific to the test sets, since we know that some false positive detections are actual galaxy clusters but we do not classify these as true positives during the analysis. We show that our model has not overfit to galaxy clusters in the training set, as we obtain a lower optimal confidence score threshold when we run the model on unseen galaxy clusters. This suggests that the features in the training set and the hyper-parameters used for our model are suitable for generalised object detection of galaxy clusters.

By applying Deep-CEE to wide-deep imaging surveys such as LSST and \textit{Euclid}, we will discover many new higher redshift and lower mass galaxy clusters. Our approach will also be a powerful tool when combined with catalogues or imaging data from other wavelengths, such as X-ray (e.g. \citealt{x_ray_emission_1}, \citealt{x_ray_emission_2}, \citealt{x_ray_emission_3}, \citealt{x_ray_emission_4}, \citealt{x_ray_emission_5}, \citealt{x_ray_emission_6}, \citealt{x_ray_emission_7} and \citealt{x_ray_emission_8}) and SZ surveys (e.g. \citealt{sz_survey}, \citealt{sz_effect_1}, \citealt{sz_effect_3}, \citealt{sz_effect_4}, \citealt{sz_effect_5}, \citealt{sz_effect_6}, \citealt{sz_effect_7}, \citealt{sz_effect_2} and \citealt{sz_effect_8}). It is hoped that the future cluster samples produced by Deep-CEE alone or in combination with other selection techniques will be well-understood and therefore applicable to constraining cosmology, as well as environmental galaxy evolution research. We will build upon this model by including methods to estimate intrinsic properties of galaxy clusters such as redshift and richness in a similar manner to George Abell many years ago.

\section*{Acknowledgements}

We would like to thank the anonymous referee for their thorough feedback which has improved the clarity of our paper. 

Funding for SDSS-III has been provided by the Alfred P. Sloan Foundation, the Participating Institutions, the National Science Foundation, and the U.S. Department of Energy Office of Science. The SDSS-III web site is \url{http://www.sdss3.org/} (\citealt{sdss_III}).

SDSS-III is managed by the Astrophysical Research Consortium for the Participating Institutions of the SDSS-III Collaboration including the University of Arizona, the Brazilian Participation Group, Brookhaven National Laboratory, Carnegie Mellon University, University of Florida, the French Participation Group, the German Participation Group, Harvard University, the Instituto de Astrofisica de Canarias, the Michigan State/Notre Dame/JINA Participation Group, Johns Hopkins University, Lawrence Berkeley National Laboratory, Max Planck Institute for Astrophysics, Max Planck Institute for Extraterrestrial Physics, New Mexico State University, New York University, Ohio State University, Pennsylvania State University, University of Portsmouth, Princeton University, the Spanish Participation Group, University of Tokyo, University of Utah, Vanderbilt University, University of Virginia, University of Washington, and Yale University. 

We acknowledge studentship funding by the Science and Technology Facilities Council (STFC), travel support provided by STFC for UK participation in LSST: All Hands Meeting through grant ST/N002512/1 and NASA’s SkyView facility located at NASA Goddard Space Flight Center.

We also would like to thank Min-Su Shu at the University of Michigan, James Schombert at the University of Oregon, Edward L. Wright at the University of California, Los Angeles \citep{cosmocalc} and Dat Tran (\url{https://github.com/datitran}) for allowing free distribution of their code.

%%%%%%%%%%%%%%%%%%%%%%%%%%%%%%%%%%%%%%%%%%%%%%%%%%


%%%%%%%%%%%%%%%%%%%%%%%%%%%%%%%%%%%%%%%%%%%%%%%%%%

%%%%%%%%%%%%%%%%% APPENDICES %%%%%%%%%%%%%%%%%%%%%

\appendix

%\clearpage
\setcounter{table}{0}
\renewcommand{\thetable}{A\arabic{table}}  

\begin{table*}
    \caption{The Total, RPN and DN loss values at different steps during the training of our model evaluated on the test set.}
    \begin{filecontents*}{All_losses.csv}
Epoch,Total loss,RPN objectness loss,RPN localization loss,DN classification loss,DN localization loss
147,0.5954,0.3825,0.0535,0.1162,0.0432
204,0.3448,0.2023,0.0538,0.0605,0.0282
313,0.3185,0.1242,0.0515,0.0794,0.0634
450,0.2841,0.0834,0.0487,0.0994,0.0526
581,0.2559,0.0694,0.0436,0.0838,0.0591
934,0.1641,0.0441,0.0380,0.0552,0.0268
1065,0.1347,0.0401,0.0302,0.0367,0.0277
1361,0.1399,0.0344,0.0283,0.0543,0.0229
1455,0.2004,0.0327,0.0268,0.1129,0.0280
1582,0.1242,0.0302,0.0279,0.0443,0.0217
1709,0.1049,0.0307,0.0256,0.0274,0.0212
1836,0.1465,0.0280,0.0261,0.0646,0.0277
1960,0.1344,0.0281,0.0301,0.0531,0.0232
2083,0.1090,0.0290,0.0246,0.0377,0.0177
2209,0.1322,0.0274,0.0237,0.0521,0.0290
2335,0.1540,0.0284,0.0256,0.0712,0.0288
2461,0.1096,0.0267,0.0247,0.0365,0.0218
2588,0.0942,0.0273,0.0245,0.0244,0.0181
2716,0.1184,0.0269,0.0354,0.0339,0.0223
2842,0.0986,0.0253,0.0235,0.0324,0.0173
2968,0.0969,0.0247,0.0222,0.0310,0.0190
3187,0.1048,0.0251,0.0275,0.0326,0.0196
3312,0.1293,0.0234,0.0233,0.0611,0.0214
3438,0.1125,0.0240,0.0224,0.0458,0.0204
3566,0.0900,0.0250,0.0232,0.0243,0.0175
3697,0.0901,0.0247,0.0213,0.0256,0.0185
3823,0.1237,0.0233,0.0238,0.0548,0.0219
3948,0.1189,0.0233,0.0229,0.0520,0.0208
4073,0.1059,0.0231,0.0208,0.0420,0.0201
4198,0.1015,0.0239,0.0217,0.0337,0.0223
4321,0.1177,0.0224,0.0204,0.0509,0.0240
4543,0.1118,0.0231,0.0221,0.0461,0.0205
4675,0.0943,0.0232,0.0215,0.0294,0.0202
4805,0.1042,0.0232,0.0264,0.0297,0.0249
4944,0.1024,0.0222,0.0194,0.0398,0.0209
5072,0.1038,0.0223,0.0201,0.0423,0.0191
5198,0.0881,0.0241,0.0205,0.0253,0.0182
5330,0.1047,0.0220,0.0202,0.0425,0.0199
5462,0.1275,0.0219,0.0241,0.0541,0.0274
5587,0.0969,0.0233,0.0220,0.0294,0.0222
5917,0.0972,0.0224,0.0265,0.0265,0.0219
6048,0.0943,0.0221,0.0184,0.0344,0.0194
6179,0.1038,0.0228,0.0228,0.0313,0.0268
6495,0.1066,0.0235,0.0231,0.0352,0.0248
6698,0.1035,0.0216,0.0184,0.0446,0.0189
6824,0.1022,0.0223,0.0190,0.0427,0.0182
6951,0.1045,0.0217,0.0187,0.0460,0.0180
7077,0.1029,0.0217,0.0188,0.0378,0.0247
7204,0.0910,0.0208,0.0180,0.0341,0.0181
7332,0.0974,0.0215,0.0176,0.0403,0.0181
7458,0.0926,0.0249,0.0211,0.0268,0.0198
\end{filecontents*}
\csvreader[
      tabular=lrrrrr,
      table head=\toprule\bfseries Step &\bfseries Total loss &\bfseries RPN objectness loss &\bfseries RPN box regression loss &\bfseries DN classification loss &\bfseries DN box regression loss\\\midrule,
      late after line=\\,
      late after last line=\\\bottomrule,
      before reading={\catcode`\#=12},after reading={\catcode`\#=6}
    ]{All_losses.csv}{1=\ItemOne,2=\ItemTwo,3=\ItemThree,4=\ItemFour,5=\ItemFive,6=\ItemSix}{\ItemOne & \ItemTwo & \ItemThree & \ItemFour & \ItemFive & \ItemSix}
    \label{tab:all_losses_table}
\end{table*}

%%%%%%%%%%%%%%%%%%%%%%%%%%%%%%%%%%%%%%%%%%%%%%%%%%

% Don't change these lines
\bsp	% typesetting comment
\label{lastpage}
\end{document}